%% file: 0-main.tex
\documentclass[lettersize,journal]{IEEEtran}
\usepackage{amsmath,amsfonts}
\usepackage{algorithm}
\usepackage{array}
\usepackage{algpseudocode}
\usepackage{graphicx}
\usepackage{float} 
\usepackage{tabularx}
\usepackage{subfig}
\usepackage{adjustbox}
\usepackage{hyphenat}
\usepackage{blindtext}
\usepackage{footnote}
\usepackage{textcomp}
\usepackage{glossaries}
\usepackage{textcomp}
\usepackage[table]{xcolor}
\usepackage{adjustbox}
\usepackage{pdftexcmds}
\usepackage{fancyvrb}
\usepackage{pgfplots}
\usepgfplotslibrary{groupplots}
\usepackage{multirow}
\usepackage{balance}
\usepackage{url} 
\usepackage[hidelinks,hyperfootnotes=true]{hyperref}

\newcommand*\circled[1]{\tikz[baseline=(char.base)]{
            \node[shape=circle,draw,inner sep=1.5pt] (char) {#1};}}
\AtBeginDocument{%
  \providecommand\BibTeX{{%
    \normalfont B\kern-0.5em{\scshape i\kern-0.25em b}\kern-0.8em\TeX}}}

\usepackage{tikz} 
\usetikzlibrary{automata} 
\usetikzlibrary{positioning} 
\usetikzlibrary{arrows} 
\tikzset{
    node distance=2.5cm, 
    every state/.append style={semithick,fill=gray!10},
    initial/.append style={initial text={}, initial distance=0.5cm},
    accepting/.append style={double=gray!10, double distance=2pt, outer sep=1pt},
    every edge/.append style={draw,->,>=stealth', auto, semithick}
}
\let\epsilon\varepsilon

\input{8.glossaries}
\begin{document}

% tdsc
%\title{Locking Down Relay and Spoofing Attacks during Concurrent Connection Establishments in 802.11ax} 
% arxiv
\title{Securing Wi-Fi 6 Connection Establishment Against Relay and Spoofing Threats}

%arxiv
\author{\IEEEauthorblockN{Naureen Hoque and Hanif Rahbari}\\
\IEEEauthorblockA{
\textit{Rochester Institute of Technology, 
NY, USA} \\
\{naureen.hoque, rahbari\}@mail.rit.edu}
}

% tdsc
% \author{Naureen Hoque, \textit{Graduate Student Member, IEEE,} and Hanif Rahbari, \textit{Member, IEEE}
% \thanks{Naureen Hoque and Hanif Rahbari are with the Golisano College of Computing and Information Sciences at the Rochester Institute of Technology, Rochester, NY, USA.}}

%\markboth{IEEE Transactions on Dependable and Secure Computing,~Vol.~x, No.~x, MONTH~202X}%{How to Use the IEEEtran \LaTeX \ Templates}

\maketitle

\begin{abstract}
Wireless local area networks remain vulnerable to attacks initiated during the connection establishment (CE) phase. Current Wi-Fi security protocols fail to fully mitigate attacks like man-in-the-middle, preamble spoofing, and relaying. To fortify the CE phase, in this paper we design a backward-compatible scheme using a digital signature interwoven into the preambles at the physical (PHY) layer with time constraints to effectively counter those attacks. This approach slices a MAC-layer signature and embeds the slices within CE frame preambles without extending frame size, allowing one or multiple stations to concurrently verify their respective APs' transmissions. The concurrent CEs are supported by enabling the stations to analyze the consistent patterns of PHY-layer headers and identify whether the received frames are the anticipated ones from the expected APs, achieving 100\% accuracy without needing to examine their MAC-layer headers. Additionally, we design and implement a fast relay attack to challenge our proposed defense and determine its effectiveness. We extend existing open-source tools to support IEEE 802.11ax to evaluate the effectiveness and practicality of our proposed scheme in a testbed consisting of USRPs, commercial APs, and Wi-Fi devices, and we show that our relay attack detection achieves 96-100\% true positive rates. Finally, end-to-end formal security analyses confirm the security and correctness of the proposed solution.
%work contributes to WLAN security by providing a robust, practical, and backwards-compatible defense against pre-authentication attacks in enterprise and public Wi-Fi networks.
\end{abstract}


\input{1.introduction}

\input{2.background}

\input{3.systhreat}
\input{4.proposed}

\input{5.evaluation}

\input{6.related}
\input{7.conclusion}

\balance
\bibliographystyle{IEEEtran}
\bibliography{IEEEabrv,9.reference.bib}

% Despite continuous growth in wireless local area network deployments, they remain vulnerable to multistage attacks initiated during connection establishments (CEs). Current security protocols fail to fully protect this phase against advanced attacks like man-in-the-middle, preamble spoofing, and relaying. To fortify this phase, we identify the underlying challenges and then design a scheme using a digital signature interwoven into the preambles at the physical (PHY) layer with time constraints. This approach slices a signature and embeds the slices within CE frame preambles without extending frame size, allowing stations to verify their respective APs' transmissions. Concurrent CEs are supported by allowing stations to analyze PHY layer header patterns to determine if a frame is from an expected AP, achieving 100% accuracy without needing to examine the MAC layer header. We design and implement a faster relay attack than existing ones to challenge and potentially bypass our proposed defense. Through experiments with USRP and Wi-Fi devices, we show that our proposed detection achieves 96-100% true positive rates. We extend open-source tools to support IEEE 802.11ax to evaluate our proposed scheme's practicality in a testbed consisting of USRP, commercial APs, and wireless devices. Finally, formal security analyses confirm the security of our proposed scheme.

\end{document}

%% file: 8.glossaries.tex
\makeglossaries

\newacronym{fcj}{FCJ}{Friendly CryptoJam}

\newacronym{phy}{PHY}{physical}

\newacronym{sr}{SR}{success rate}

\newacronym{awgn}{AWGN}{additive white Gaussian noise} 

\newacronym{ce}{CE}{connection establishment 
} 
\newacronym{fj}{FJ}{friendly jamming}

\newacronym{fc}{FC}{fully connected}

\newacronym{tcm}{TCM}{Trellis-coded modulation}

\newacronym{cnn}{CNN}{convolutional neural network}

\newacronym{dl}{DL}{deep learning}

\newacronym{snr}{SNR}{signal-to-noise ratio}

\newacronym{rf}{RF}{radio frequency}

\newacronym{wlan}{WLAN}{wireless local area network}

\newacronym{ap}{AP}{access point}

\newacronym{mitm}{MitM}{man-in-the-middle}

\newacronym{krack}{KRACK}{key reinstallation attack}

\newacronym{csa}{CSA}{channel switching announcement}

\newacronym{dos}{DoS}{denial-of-service}

\newacronym{pmk}{PMK}{pairwise master key}

\newacronym{eap}{EAP}{extensible authentication protocol}

\newacronym{msk}{MSK}{master session key}

\newacronym{ptk}{PTK}{pairwise transient key}

\newacronym{vht}{VHT}{very high throughput}

\newacronym{wba}{WBA}{Wireless Broadband Alliance}

\newacronym{epmod}{eP-Mod}{extensible preamble modulation}

\newacronym{wani}{WANI}{Wi-Fi Access Network Interface}

\newacronym{ocv}{OCV}{operating channel validation}

\newacronym{oci}{OCI}{operating channel information}

\newacronym{tsf}{TSF}{timing synchronization function}

\newacronym{mc}{MC}{model checker}

\newacronym{cpv}{CPV}{cryptographic protocol verifier}

\newacronym{fsm}{FSM}{finite state machine}

\newacronym{ctl}{CTL}{computational tree logic}

\newacronym{ltl}{LTL}{linear tree logic}

\newacronym{hmac}{HMAC}{hash-based message authentication code}

\newacronym{mac}{MAC}{message authentication code}

\newacronym{ber}{BER}{bit error rate}

\newacronym{nist}{NIST}{National Institute of Standards and Technology}

\newacronym{cfo}{CFO}{carrier frequency offset}

\newacronym{rss}{RSS}{received signal strength}

\newacronym{ocvc}{OCVC}{oerating channel validation capable}

\newacronym{sa}{SA}{security association}

\newacronym{mfp}{MFP}{management frame protection}

\newacronym{tpr}{TPR}{true positive rate}

\newacronym{fnr}{FNR}{false negative rate}

\newacronym{fpr}{FPR}{false positive rate}

\newacronym{roc}{ROC}{receiver operating characteristic}

\newacronym{auc}{AUC}{area under ROC curve}

\newacronym{pca}{PCA}{principal component analysis}

%% file: 1.introduction.tex
\section{Introduction}\label{introduction}

The adoption of \gls*{wlan} has experienced \textit{exponential} growth since 2020~\cite{wbareport2023}. The Wi-Fi 6/6E market reached \$2.3 billion within just a few years, the Enterprise segment generated \$3.5 trillion in economic value in 2021 alone~\cite{wbareport2022}, and 802.1X-based OpenRoaming™ for public networks surged to $3.5$ million hotspots in densely populated areas like airports and stadiums by 2023~\cite{wbareport2023}. However, this growing wireless infrastructure has been vulnerable to various forms of multi-stage attacks, often initiated by exploiting the unprotected \textit{\gls{ce} phase}~\cite{vanhoef2017key,vanhoef2020dragonblood,vanhoef2021fragment,gvozdenovic2020truncate}. The latest WPA3 specification for Wi-Fi security, including the IEEE 802.11w amendment for management frames, is still designed to secure frames at the MAC layer only after successful mutual authentication, known as the four-way handshake, with the exception of optional operating channel and beacon integrity protections. This leaves the preamble (at the \gls*{phy} layer) and payload of the management frames before that point largely unprotected. 

%These attacks eventually can starve users of access to idle channels~\cite{gvozdenovic2020truncate,zhang2021preamble}, disrupt frame detection and alter data~\cite{zhang2021preamble}, decrypt (data) packets, relay or replay frames, or in certain cases, retrieve the authentication key~\cite{vanhoef2017key,vanhoef2020dragonblood}. 

A pre-authentication\footnote{We use the terms connection establishment and pre-authentication phase interchangeably in this article.} exploit can enable an adversary to spoof a frame preamble, or establish a \gls*{mitm} to selectively relay, block, or replay frames. For example, a forged preamble can be leveraged to starve a receiver of access to idle channels by making it await a non-existent payload~\cite{gvozdenovic2020truncate,zhang2021preamble}. Likewise, by obscuring the preamble signal, the adversary can disrupt frame detection at the receiver, leading to incorrect frame decoding~\cite{gvozdenovic2020truncate,lapan2016physical}. In addition, offering a higher signal strength on a different channel (measured using the preamble), abusing the unprotected \gls*{csa} element, or jamming the channel of the real \gls*{ap} during the \gls{ce} are common methods to launch a multi-channel \gls*{mitm} attack~\cite{thankappan2022multi}.
%vanhoef2014advanced,vanhoef2017key,vanhoef2020dragonblood, vanhoef2021fragment}.
An \gls{mitm} position, in turn, can be used to launch more advanced attacks including decrypting (data) packets~\cite{schepers2023framing}, altering beacon frames~\cite{vanhoef2014advanced}, retrieving the authentication key in certain cases~\cite{vanhoef2017key,vanhoef2020dragonblood}, or, as we experimentally study in this paper, launching a selective relay attack.

%Spoofing a frame preamble or deceiving a station into connecting to a \gls*{mitm} are two such attacks. The frame preamble, which is used at the \gls*{phy} layer to indicate the start and duration of a frame, can be leveraged to starve a receiver by using a forged preamble to make it await a non-existent payload~\cite{gvozdenovic2020truncate,zhang2021preamble}. Moreover, by obscuring the preamble signal start, the attacker can complicate frame detection for the receiver, leading to incorrect frame decoding~\cite{gvozdenovic2020truncate}. 

%To protect against preamble spoofing attacks, Zhang \textit{et al.} recently proposed customizing preambles using existing group keys and timestamps but not until after \gls*{ce}a relay attack faster than existing ones is designed and implemented to challenge our proposed defense and determine whether it can circumvent our proposed defense.~\cite{zhang2024preamble}. However, as timestamps are not protected frame elements, adversaries can easily spoof them, thereby bypassing the defense. 
To protect against preamble spoofing attacks, Zhang \textit{et~al.} have proposed customizing the traditionally fixed and known Wi-Fi preambles using timestamps and the same group key as that for protecting the beacons, but not until after \gls*{ce} phase~\cite{zhang2024preamble}. However, since timestamps are not protected frame elements, adversaries can easily spoof them, thereby bypassing their defense. Additionally, even if this defense is extended to protect the CE phase, it would be unable to detect if the adversary relays a frame—i.e., captures and forwards a frame from an AP to a station, or vice versa—by replaying a legitimate customized preamble but alters other parts of the frame. 
To protect the operating channels to prevent multi-channel \gls*{mitm} attacks
the IEEE 802.11-2020 standard introduced the \gls*{ocv} to safeguard the \gls*{csa} field at the MAC layer. However, \gls*{ocv} cannot protect against other forms of multi-channel \gls*{mitm} attacks mentioned above and other pre-authentication threats like preamble spoofing. This underscores the need to also verify APs' transmissions at the radio signal level (\gls*{phy} layer) to comprehensively address and fully protect against all pre-authentication threats.

%To defend against pre-authentication attacks, the IEEE 802.11-2020 standard added the \gls*{ocv} defense mechanism to protect the \gls*{csa} element field at the MAC layer. However, this technique solely counters \gls*{csa}-based multi-channel \gls*{mitm} attacks~\cite[§,12.2]{80211}; it cannot even defend against jamming-based multi-channel \gls*{mitm} (relay) attacks. This narrowly scoped amendment was made without considering a wider range of other pre\hyp{}authentication attacks, e.g., preamble spoofing. To protect the \gls*{ce} phase, it is critical to authenticate an \gls*{ap} at the radio signal level to jointly counter these attacks since the MAC layer is oblivious to the aforementioned spoofing and relay attacks both at the \gls*{phy} and MAC layer.

We proposed a preliminary technique in~\cite{hoque2023countering} to protect the \gls*{ce} phase against relay and preamble spoofing attacks by effectively interweaving AP authentication across the \gls*{phy} and MAC layers. It integrates a practical digital signature scheme from the MAC layer with a time constraint mechanism for the \gls*{ce} phase to verify the transmissions at the \gls*{phy} layer under the IEEE 802.11ac standard. Specifically, in our scheme, a legitimate AP first signs its MAC address and a timestamp, and then slices this signature and embeds one slice in each unicast pre\hyp{}authentication frame's preamble, \textit{chaining} together all these frames reinforced with timing constraints. This temporal control effectively blocks attempts to relay or replay authentic preambles as a means to, e.g., manipulate other frame elements. Together with the signature, they offer a robust defense mechanism for enterprise and 802.1X-based Wi-Fi networks (including Wi-Fi Certified Passpoint) against relay and spoofing attacks. It is a backward-compatible solution that avoids extending frame size or transmitting additional frames thanks to a specific preamble-embedding technique~\cite{zhang2021adaptive} we use, thereby adding negligible communication overhead each time a station needs to verify an AP.

%It leverages the existing trust infrastructure in such networks, specifically the authentication server that itself is verified using the 802.1X framework, by requiring it to generate a public-private key pair for each AP. 
%We use a robust %\gls*{nist}-approved signature that is short enough to be split and distributed across the unicast pre\hyp{}authentication frames' (sent by an AP) preambles when using a specific preamble-embedding technique~\cite{zhang2021adaptive}, one that makes our solution backward-compatible while adding negligible communication overhead each time a station needs to verify an AP, thereby avoiding extending frame size or transmitting additional frames. 

However, we did not address two critical issues in our preliminary work: (a) our relay attack detection mechanism was evaluated only through simulations and assuming ideal channel scenarios; and (b) we considered only a single station trying to connect to an AP—a simplified context for our proposed defense. In this paper, we first design a USRP transceiver that relays the APs' frames with minimal latency, the first of its kind to our best knowledge, for experimentally assessing our proposed time-bound approach under worst-case scenarios, where the attacker attempts to immediately relay frames as soon as they are captured without any alteration. The existing works~\cite{vanhoef2014advanced, schepers2023framing} demonstrate multi-channel MitM attacks requiring the overhead of spoofing or alteration of select frame 
%the CSA and timestamp 
elements at the MAC layer. 
%—allowing the use of any previously captured beacon frame. %—our relay attack does not necessitate any changes to the frame.  
We argue that our relay attack design allows a stronger threat model to cover more complex relay attacks, %that 
which require altering frame elements that introduce additional delays. 
A pivotal part of this contribution is using the high-speed, performance-optimized USRP X310 with a PCIe interface enabling rapid relay attacks to challenge the proposed time-bound technique.

%It aims to relay a frame as soon as it is received within the expected timeout period of the victim station, ensuring rapid execution. 
%We experimentally demonstrate the attack using commercial Wi-Fi devices to evaluate the detection performance, achieving 96-100\% true positive rates. 

The second challenge we address is concurrent \glspl*{ce} involving multiple APs and stations. It is crucial for a station in our approach to accurately determine at the PHY layer which frames contain signature slices it expects among a diverse set of frame types, such as beacons, acknowledgment frames, and pre-authentication frames intended for other stations, without needing to inspect the frame type in the MAC layer header. To address this challenge, in this paper we meticulously analyze the SIG (Signal) field within the PHY layer header, searching for specific patterns that distinguish between different frame types. Given that these are pre-authentication frames, which typically exhibit less variation compared to data frames, they often demonstrate a degree of a consistent pattern in terms of frame size, duration, and rate—details that are in the SIG field. By leveraging this consistent pattern, we demonstrate that a station can indeed identify, solely at the PHY layer and without delving into the MAC layer, whether a received frame is a beacon, acknowledgment, or an anticipated \gls*{ce} frame. Employing \gls{pca}—a statistical method that identifies directions of maximum data variance—we show that a station can differentiate among multiple APs and distinct frame types with 100\% accuracy.

To evaluate the performance of our work under the latest Wi-Fi standard 802.11ax, we conduct extensive simulations and real-world experiments. Our new simulations incorporate various channel models and a realistic Wi-Fi environment, enhancing beyond the scope of our preliminary work. Furthermore, to do the over-the-air experiments with commercial AP-USRP testbed, in this paper, we extend the \textit{gr-ieee802-11}~\cite{bloessl2018performance} library to support 802.11ax, advancing from our preliminary work on 802.11ac~\cite{hoque2023countering}, and preamble modification. We have also verified that the current draft of IEEE 802.11be (the upcoming standard--Wi-Fi 7), does not amend any CE components~\cite{rodriguez2021ieee80211be}. 

%Additionally, to experiment with our scheme under a more realistic setup, we first extend the \textit{gr-ieee802-11}~\cite{bloessl2018performance} library to support IEEE 802.11ax ($20$ MHz bandwidth), advancing from our preliminary work on IEEE 802.11ac~\cite{hoque2023countering}, and preamble modification. %Commercially available APs restrict modifying the preamble while open-source full-stack implementations of Wi-Fi (e.g., Openwifi~\cite{openwifigithub}) currently do not support the IEEE 802.11ac/ax protocol or preamble modification. Then we implement \textit{eP-Mod}, and finally experimentally evaluate our proposed scheme. 

%To demonstrate the effectiveness of our relay detection mechanism, we design a USRP transceiver that captures frames from APs and relays them with minimal latency, the first of its kind to our best knowledge, for assessing our proposed time-bound approach under worst-case scenarios; and conducting thorough experiments using commercial Wi-Fi devices to evaluate the relay detection performance, achieving an impressive 97.5\% accuracy and 96-100\% true positive rates in identifying relay attacks. A pivotal part of this contribution is the use of the high-speed, performance-optimized Ettus Research USRP X310 enabling rapid relay attacks to challenge the time-bound technique. This thorough testing validates our approach's practical applicability and effectiveness in real-world scenarios. 

Furthermore, we perform a comprehensive end-to-end formal security analysis of the proposed technique, including the time-bound method.  Also, in this paper, we provide a new formal, end-to-end security verification employing a symmetric key approach (i.e., HMAC) to demonstrate the robustness of our method, affirming its effectiveness across both symmetric and public key approaches. Upon acceptance, we will make all code available as open-source. 

\textit{Contributions--} Our main contributions are as follows:

\begin{enumerate}
    \item We design a novel defense mechanism to protect the Wi-Fi CE phase from relay and spoofing attacks by effectively authenticating at both the PHY and MAC layers the frames received at the stations, supporting complex real-world scenarios of concurrent connection attempts involving multiple nearby APs and stations and diverse frame reception under IEEE 802.11ax (Wi-Fi 6).

    \item We design and demonstrate a fast relay attack with minimal latency using a USRP transceiver, the first of its kind to our best knowledge, and leveraging it to test our time-bound approach for relay detection under worst-case scenarios, achieving $96$-$100\%$ true positive rates in attack detection using commercial Wi-Fi devices. 
    
    \item Our extensive simulations and experiments show that (i) our scheme simulated across different channel models demonstrates a success rate in signature transmission via preambles of $98$-$100\%$ at $5$\,dB SNR; (ii) experiments with various commercial Wi-Fi devices show a $100\%$ accuracy in identifying all five tested APs at the and diverse frame types at the PHY layer; (iii) experiments on a commercial AP-USRP testbed under 802.11ax setups confirmed the efficiency and practicality, adding only an average delay of $2.19\%$ to total CE time.

    \item We formally model and prove the correctness of our technique including the time-bound method against various attacks using a model checker (MC), and then verify its end-to-end integrity and authenticity using a cryptographic protocol verifier (CPV)- considering both symmetric and asymmetric approaches.
\end{enumerate}

\textit{Paper Organization-- }The remainder of this paper is organized as follows. We first provide the necessary background and our system and adversary models in Sections~\ref{background} and~\ref{model}, respectively. Our proposed scheme is described in Section~\ref{proposed}. We present our evaluation results in Section~\ref{evaluation} before reviewing related work in Section~\ref{relatedwork}. We conclude in Section~\ref{conclusion}.

%% file: 2.background.tex
\section{Preliminaries}\label{background}

We briefly review the Wi-Fi connection establishment phase, relevant Wi-Fi frame elements, pre\hyp{}authentication relay and spoofing attacks at the PHY-layer, and preamble embedding. 

\subsection{Secure Connection Establishment}\label{connect}
\begin{figure}[t]
    \centering
    \includegraphics[scale=0.2]{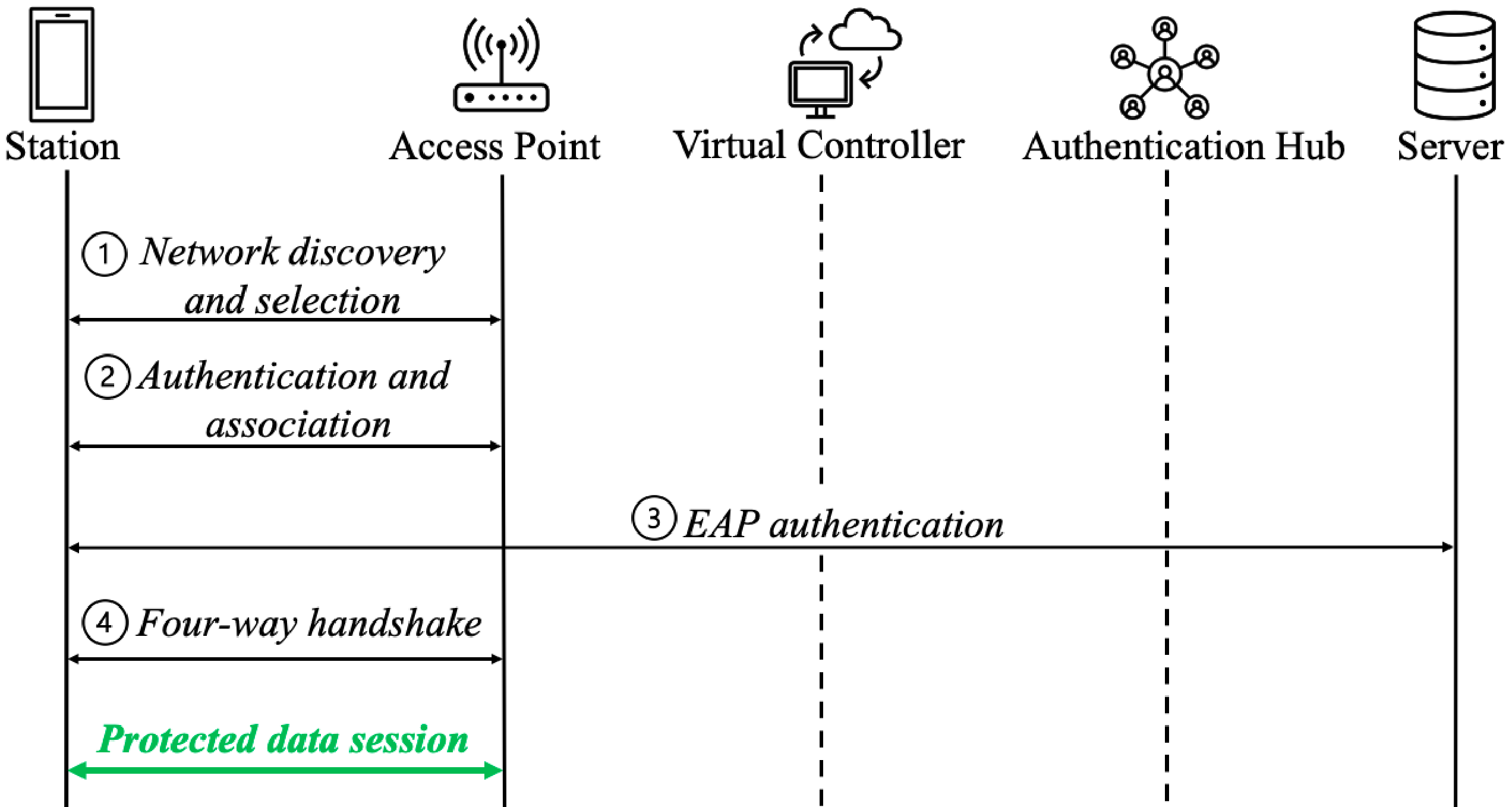}
    \caption{Wi-Fi connection establishment (virtual controller and authentication hub are present in Passpoint® networks).}
    \label{wifi-con-est}
\end{figure}

Enterprise Wi-Fi networks follow a structured process for secure connection establishment that involves an authentication server, \gls*{ap}, and stations. The server, verified via a certificate authority, manages user credentials and is responsible for generating a \gls{pmk} for each AP-station pair. The architecture and the connection establishment in Passpoint® (a.k.a. Hotspot 2.0) are similar to those of the Wi-Fi enterprise (see~\figurename~\ref{wifi-con-est}). This is also true for OpenRoaming™, eduroam, and any other 802.1X-enabled public Wi-Fi networks (they are in contrast to traditional open Wi-Fi networks that offer no security at all, or Enhanced Open™ that provide only unauthenticated data encryption~\cite{owe})

The connection establishment begins with network discovery and selection, where stations select APs based on the highest received \textit{signal strength} measured using those frames' training (preamble) signals~\cite[§\,17.3.12]{80211}. This \gls{ce} phase involves either active or passive scanning, with the station and AP exchanging a series of unprotected management frames.

Following the initial discovery, the station engages in authentication and association with its chosen AP. Then, the \gls*{eap} method along with 802.1X is used in the enterprise and public modes to facilitate mutual authentication between the station and the server. Commonly used \gls*{eap} methods include EAP-PEAP and EAP-SIM, which securely transmit authentication information through the AP, culminating in the derivation of the \gls*{pmk}. The \gls*{pmk} is derived using a \gls{msk}, which is sent to the station via the AP using one of the \textit{EAP} frames. 

The final phase involves a four-way handshake between the AP and the station using the \gls*{pmk} to establish a \gls{ptk} for securing all further communications beyond \gls*{ce} phase. Excluding beacons and potential retransmissions, this process necessitates the transmission of at least 13 and 15 unique unprotected management frames for EAP-SIM and EAP-PEAP methods, respectively.

\subsection{Relevant Wi-Fi Frame Elements}
%In this section, we will provide the relevant frame properties of Wi-Fi to follow the rest of our paper. 

\subsubsection{Frame Preamble}
Every Wi-Fi frame is prepended at the \gls*{phy} layer by a training signal that is used by the receiver to perform certain PHY-layer functions, including frame detection, received signal strength estimation, and synchronization~\cite[§\,17.3]{80211}. Those signals together with the SIG field (used to indicate the frame duration, among other \gls*{phy}-layer information, then form a preamble.
%The frame preamble at the PHY layer serves to indicate the start and duration of a frame, which is achieved through training signals and the SIG field, respectively. Specifically, the training signals are utilized by the receiver for various PHY-layer functions, such as frame detection, received signal strength estimation, and synchronization.\cite[§\,17.3]{80211}. 

\subsubsection{Channel Switch Announcement Element}\label{sec-csa}

To change the operating channel in the middle of a connection establishment (e.g., when the current channel has a poor quality or has to be vacated for radar in proximity~\cite[§\,11.8]{80211}), an AP uses the \gls{csa} element, which can be sent within a beacon \textit{anytime} during this phase, to advertise when it intends to switch to a specific channel. This element can also be part of an action frame (a type of management frame to already-associated stations) or a probe response. % which are unprotected, %as these are pre\hyp{}authentication frames and so can be spoofed by an adversary.
%An \gls{ap} may change its operating channel in the middle of the connection establishment for various reasons, such as, the current operating channel is congested, or has poor quality, or needs to be vacated for a radar (specific to 5 GHz). It uses the \gls{csa} element, which can be sent within a beacon \textit{anytime} during this phase, to advertise when it intends to switch the channel. %Dynamic Frequency Selection (DFS) function in Wi-Fi enables the option for an \gls{ap} to use 5 GHz frequencies that are reserved for weather radars. When an \gls{ap} enables this function, it uses the Channel Availability Check (CAC) to confirm if any of the DFS frequencies is used by a radar in proximity. %If a radar is detected using a specific channel, then the \gls{ap} excludes that channel from the list of available channels and after a certain period, the \gls{ap} rechecks if that channel is available or not. 
%If the \gls{ap} needs to switch its active operating channel to make that available for a radar, then it uses the Channel Switch Announcement (CSA) element to advertise a new channel number to all its associated stations. 

\subsubsection{Frame Sequence Number, Timeout, Retransmission}
To account for possible frame transmission failures, the standard defines a \textit{timeout} interval for every frame and, in turn, allows retransmission of a lost or corrupted frame after that~\cite[§\,9-10]{80211}. This interval includes transmission time, propagation and processing delays, inter-frame space and slot time, etc. %, and ranges from $\sim0.045-20$\,ms depending on the frame type (e.g., management, control). 
Multiple retransmissions are allowed within a predefined \emph{retry limit} until the frame is successfully received~\cite[§\,10]{80211}. The sequence number of a frame is unique but remains constant in all retransmissions~\cite[§\,9]{80211}.

\subsubsection{Device Location Element}\label{devicelocation} This frame element includes the location information (latitude, longitude, altitude, etc.) and can be used by a station/AP to announce their location to others~\cite[§\,9.4]{80211}.

\subsection{Spoofing/Relay Attacks in Wi-Fi}
\textit{Spoofing Preambles---} The preamble is not protected by the existing security protocols. Therefore, an adversary can spoof it to launch advanced multi-stage attacks---see Table~\ref{root-cause}. For example, it can force a station into silence (not able to send/receive) by sending a fake preamble but not its expected subsequent payload, effectively starving any receiver that has detected that preamble of channel access~\cite{zhang2021preamble}.

\textit{Relaying Frame Elements---} In Wi-Fi, any \gls{mitm} (or relay) must be a \textit{multi-channel} one because an adversary cannot use a fake AP MAC address (the station can detect this during four-way handshake) and attempting to use the same MAC address as a real AP on that AP's channel will be easily detected by the legitimate AP. Thus, setting up with a real AP MAC address on a different channel is the attacker's only option (unless it uses a directional antenna to be able to relay while being on the same channel). In a CSA-based MitM, the rogue AP sends a spoofed \gls*{csa} element to the station to make it switch its channel~\cite{vanhoef2017key}. Table~\ref{root-cause} lists of a few advanced attacks that are initiated by a CSA-based \gls*{mitm} attack. Likewise, a jamming-based \gls*{mitm} involves jamming the real AP and having the station join a rogue AP on a different channel, as does offering a higher signal strength on a different channel without jamming. %, the real AP is jammed and the station if orchestrated to join a rogue AP. A rogue AP offers a stronger signal on the other channel in a higher signal strength-based \gls*{mitm}. Table~\ref{root-cause} lists a few of these attacks. % a \gls*{mitm} launches by relaying, delaying, or blocking the frames. % (see Section~\ref{preamble}), with a station choosing to connect to it~\cite{vanhoef2020dragonblood}. 

\textit{Validating Operating Channel---} \gls{ocv} can protect \gls*{csa} elements~\cite[§\,12.2]{80211} by mandating an authenticated \gls*{oci} element in each frame, preventing \gls*{csa}-based \gls*{mitm} attacks~\cite{vanhoef2018operating}.
This technique adds $7\%$ latency to the existing connection establishment because of the extra frames exchanged in each channel switch~\cite{hoque2022systematically}. Also, this technique cannot prevent a non-CSA \gls*{mitm} (relay) attack or a preamble-based spoofing one.

\begin{table}[t]
    \caption{Recent multi-stage attacks on Wi-Fi networks that are initiated by spoofing unprotected frame elements.}  \label{root-cause}
    \centering
    \begin{tabular}{|m{0.065\textwidth}|p{0.36\textwidth}|}
        \hline
        Element & {\centering Attacks} \\
        \hline
        Preamble & 
        \gls*{mitm}~\cite{vanhoef2014advanced}, channel silencing~\cite{zhang2021preamble}, TaP attack~\cite{gvozdenovic2020truncate}, data alteration~\cite{zhang2021preamble}, frame detection attack~\cite{zhang2021preamble}\\
        \hline
        CSA & KRACK~\cite{vanhoef2017key}, Dragonblood~\cite{vanhoef2020dragonblood}, FragAttack~\cite{vanhoef2021fragment}, \gls*{mitm}~\cite{vanhoef2014advanced},  group-key attack~\cite{vanhoef2016predicting}, FramingFrames~\cite{schepers2023framing}\\
        \hline
    \end{tabular}
\end{table}

\subsection{Embedding Bits in the Preamble Signal}\label{preamble}

%\textit{Bit Embedding in the Preamble---}
Using the \gls{epmod} technique~\cite{zhang2021adaptive}, user-defined bits can be embedded in the preamble\footnote{Henceforth, \textit{preamble} refers specifically to the training signal in this paper.} in a backward-compatible way. Specifically, the preamble signal in 802.11ac Wi-Fi systems is shown to be able to reliably contain up to $20$ bits per frame with a $40\,$MHz channel and achieve the same or even better \gls*{ber} performance of the BPSK modulation scheme~\cite{zhang2021adaptive}. More bits can be embedded with more spatial multiplexing or with less stringent \gls*{ber} performance even under noisy channel conditions, as extensively studied in~\cite{zhang2021adaptive}. %The efficiency, feasibility, and robustness of this bit embedding technique in preamble is demonstrated in~\cite{zhang2021adaptive} with $90\%$ preamble decode ratio, while maintaining comparable frame detection, synchronization, and payload \gls*{ber} performance as the standardized preamble.

%% file: 3.systhreat.tex
\section{System \& Adversary Model}\label{model}

\textit{System Model---}We consider an IEEE 802.11ax Wi-Fi network configured with WPA3 (which enables .11w by default) in enterprise mode or IEEE 802.1X-based public mode (e.g., Passpoint®, OpenRoaming™). Alternatively, we consider any Wi-Fi network that relies on a \emph{trusted} authentication server and an EAP method to issue a valid public-private key pair for an AP and communicate the public key to the stations. We consider that at least three channels are available in the system. We further assume a legitimate AP with one or possibly two transmit antennas, meaning that an 802.11ax frame preamble from the AP can highly reliably embed up to $20$ user-defined bits over a $20\,$MHz channel using the \gls{epmod} technique~\cite{zhang2021adaptive}. Additionally, while an AP is in a connection establishment stage with one station, it can continue broadcasting periodic beacons and connect with other stations. All stations use the beacon's timestamp or the \gls*{tsf} to synchronize with the AP~\cite[§\,9.4.1.10]{80211}.

\textit{Adversary Model---}We consider the de facto adversary model in network security systems, Dolev-Yao~\cite{dolev1983onthe,hussain2018lteinspector, hussain20195greasoner}. The adversary can eavesdrop, jam, replay, relay, and modify legitimate pre\hyp{}authentication frames or their preambles, or inject new ones, but cannot decrypt the communications between the AP and the server. It has the following features: (i) The adversary has unlimited resources to create several fake \glspl*{ap} with its desired MAC address(es). Both \glspl*{ap} (real and fake) can be active at the same time, but on different channels (so the adversary evades detection). It can relay on the same channel only using a directional antenna. (ii) The rogue \gls*{ap} cannot be a part of the trusted server's network (i.e., not an insider). Also, it cannot physically tamper with a real \gls*{ap} (or station). (iii) The adversary does not have any access to the real \gls*{ap}'s private key. Likewise, the user credentials (identity and password) of a station are not available to the adversary. (iv) The embedded preamble bits are visible to the adversary because they are not encrypted. Having said that, the adversary does not and cannot have any access/knowledge of the preamble bits of a frame that an AP has not transmitted yet.

\textit{Goal---}The adversary's main goal is to relay, alter, or spoof pre\hyp{}authentication management frame(s) or signals at the PHY/MAC layers to launch an attack (e.g., starvation or multi-channel MitM). 

%% file: 4.proposed.tex
\section{Proposed Verification Techniques}\label{proposed}

We propose that the trusted authentication server in enterprise/ Passpoint \glspl*{wlan} provides a station, at the time of each \gls*{ce}, with either a symmetric or public key of the AP. Our design can accommodate each of these alternatives with different levels of scalability, granularity, and security. We start by comparing them in terms of their real-world deployment.

\subsection{Symmetric vs. Asymmetric Key Approach}\label{symasym}

A symmetric key-based verification function, such as \gls*{hmac}, is generally faster than a digital signature in terms of generation and verification. It also allows the verifier to compute the same \gls*{hmac} once it receives the message, a feature that can further enhance the security of our design (see Section~\ref{designdetails}). However, symmetric key-based approaches encounter scalability issues in practice, e.g., in airports and shopping malls. If a single symmetric key is shared among all the stations, it will be easier for an adversary to obtain that key. Alternatively, considering one such key per each AP-station pair will not be scalable due to the storage and key management complexity~\cite{hussain2019insecure}. For example, the authentication server in an international airport would need to frequently generate and maintain thousands of AP-station keys, while most of the stations (i.e., passengers) may not return to that airport ever again. Setting an expiration time for each key, after which the server and APs delete it, can solve this issue; however, it introduces an additional layer of key maintenance and management.

With an asymmetric key approach, however, the server would need to generate and maintain only one key pair (public and private) per AP. Although this makes it a more scalable approach for large-scale (enterprise/public) networks, digital signature functions (generation and verification) are slower than those of \gls*{hmac} solutions. Therefore, in scenarios in which the network size is relatively small, e.g., in an enterprise with recurring users/employees, the benefits of an HMAC-based solution with key expiration will outweigh those of an asymmetric key one. 
Without loss of generality, we explain our design assuming an asymmetric approach for both kinds of networks (enterprise and public) and discuss the advantages of using \gls{hmac} instead of digital signature only when relevant. 

\subsection{Proposed (Asymmetric Key) Scheme}\label{over}

\begin{figure}[t]
    \centering
    \includegraphics[scale=0.2]{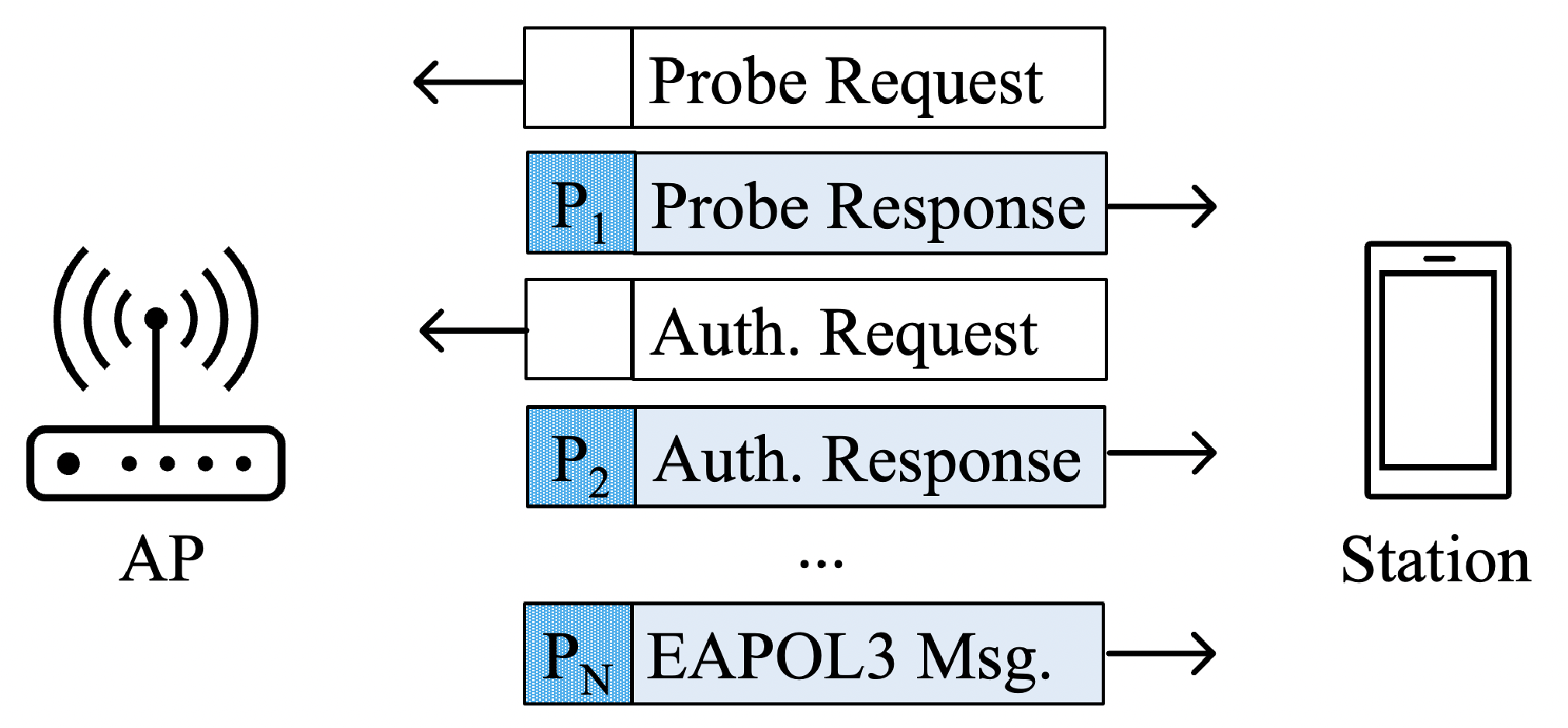}
    \caption{Each CE frame from the AP carries one signature slice embedded in the preamble $P_n$.}\label{generate}
\end{figure}

To prevent spoofing and relay attacks during the \gls*{ce} phase, the AP needs to protect the pre\hyp{}authentication frames it sends to a station. To that end, we propose that an AP generates only one signature to protect all the unprotected unicast management frames it sends to a given station (we discuss alternative methods, such as one signature per frame or only for the last frame, and their limitations in Section~\ref{limitation}). This signature should be generated at the beginning of the \gls*{ce} process and interwoven into the \gls{phy} layer preamble. We build upon the generic \gls*{epmod} communication technique described in Section~\ref{preamble} to embed slices of that signature in the preamble of those frames. Therefore, each frame carries one piece of the signature (as shown in~\figurename~\ref{generate}) to protect its preamble signal while the signature itself should be short enough that its slices can be communicated reliably (e.g., considering frame retransmissions) using the preambles. The chain of slices is further tightened using (1) the unique \textit{sequence number} of a frame to detect frame insertion and preamble replay attacks, and (2) \textit{time constraints} we impose on individual pieces of this chain to counter relaying a protected preamble when used to spoof other non-cryptographically protected content in that frame. We cryptographically protect the channel number and the sequence number of only the last frame, which helps to devise our mechanism for detecting frame insertions and tracking valid and invalid channel switches in this phase. 

Our time-bounding and signature/HMAC mechanisms are both essential for comprehensive protection. While the signature/HMAC ensures the authenticity and integrity of the MAC address and timestamp, it alone cannot thwart relay attacks or guarantee the sequence's correctness. Conversely, time bounding alone cannot assure the integrity of the operating channel number and timestamp. Using them in combination ensures both timestamp integrity (which is also critical for the time-bound solution) and protects against relay and spoofing by securing the sequence number.

We note that a symmetric-key approach can further enable checking the slice upon the arrival of each frame, a stronger measure to mitigate several PHY-layer attacks like preamble spoofing (see Section~\ref{designdetails} for details). However, both symmetric and asymmetric methods require waiting for the last frame to reconstruct the signature/HMAC for \gls*{ap} verification.

\begin{algorithm}[ht]
\caption{}
\begin{algorithmic}[1]
\Procedure{GenerateSignature}{$m, \psi_{\text{AP}}$}
    \State $\mathcal{S} \gets \text{Sign}(m, \psi_{\text{AP}})$
    \State $\mathcal{E} \gets \text{Encode}(\mathcal{S})$
    \State \textbf{return} $\mathcal{E}$
\EndProcedure
\Procedure{SliceAndEmbed}{$\mathcal{E}, \mathcal{N}$}
    \For{$i = 1$ to $\mathcal{N}$}
        \State $\mathcal{E}_i \gets \text{Slice}(\mathcal{E}, i)$
        \If{$i = \mathcal{N}$}
            \State $\mathcal{E}_i \gets \mathcal{E}_i \oplus \text{Hash}(\phi_{Ch}, \phi_{Sq})_{\text{PTK}}$
        \EndIf
        \State Embed $\mathcal{E}_i$ in preamble of frame $i$
    \EndFor
\EndProcedure
\Procedure{Verify}{$\mathcal{E}, \kappa_{\text{AP}}, t_{in}$}
    \State $\mathcal{S}' \gets \text{Empty String}$
    \For{$i = 1$ to $\mathcal{N}$}
        \If{$f_{P} = f_{s}$}
            \If{Received within $t_{in}$}
                \State $\mathcal{S}' \gets \mathcal{S}' \, || \, \mathcal{E}_i$
            \EndIf
        \EndIf
    \EndFor
    \State $\mathcal{S} \gets \text{Decode}(\mathcal{S}')$
    \State \textbf{if} $(\mathcal{S}, \kappa_{\text{AP}}), \phi_{Ch}, \phi_{Sq}$ are \textbf{correct} \textbf{then}
        \State \hspace{\algorithmicindent} \textbf{return} \textit{Success}
    \State \textbf{else}
        \State \hspace{\algorithmicindent} \textbf{return} \textit{Failure}
\EndProcedure

\end{algorithmic}
\end{algorithm}

\subsubsection*{Proposed Algorithm}

We leverage the existing trusted server to generate and maintain public and private keys for each AP and securely share them with individual APs. Additionally, along with the \gls*{msk}, the server securely sends the AP’s public key and device location to the station during the EAP process. The proposed technique is outlined in Algorithm 1 and described as follows: key notations and design details are provided in Table~\ref{notations} and Section~\ref{design}.

\begin{itemize}%[noitemsep,topsep=0pt,leftmargin=4mm]
    \item \textsc{GenerateSignature}-- Once an AP receives the first pre\hyp{}authentication frame from a station, it generates a signature $\mathcal{S}$ over a message $m$ using its private key $\psi_{\text{AP}}$. The signature is then encoded using a channel coding scheme to increase its robustness to communication errors.
    \item \textsc{SliceAndEmbed}-- The encoded signature $\mathcal{E}$ is sliced into $\mathcal{N}$ pieces, where $\mathcal{N}$ is the total number of pre\hyp{}authentication management frames ($\mathcal{N} \in \{13, 14, 15\}$---see Section~\ref{background}) that the AP will send to the station. For the last pre\hyp{}authentication frame to be sent by the AP (3rd EAPOL message---EAPOL3), the final signature slice is XORed with the cryptographically hashed operating channel number, and that frame's sequence number using the \gls{ptk} before it embeds the bits in the preamble.
    \item \textsc{Verify}-- The station first verifies that the received frame is indeed a CE frame containing a signature slice in its preamble from the expected AP (see section~\ref{concurrent} for verification details). Additionally, it checks whether the frames arrive within the specified time constraint $t_{in}$ (details in Section~\ref{timeboundsection}). Upon receiving the final signature slice, the station first verifies the frame's sequence number and the AP's operating channel against the cryptographically hashed values. Then, by assembling all the slices, the station decodes $\mathcal{E}$ into $\mathcal{S}$. If the signature is successfully verified as well, the station sends its final EAPOL message, thereby establishing a secured data session with the AP. If any of the verification fails, the station disassociates from the AP.
\end{itemize}

The algorithm's complexity primarily relies on cryptographic operations and processing frame slices. Signature generation and verification are standard cryptographic tasks, typically with $O(1)$ complexity per operation, assuming constant-time algorithms. Thus, the \textsc{GenerateSignature} procedure, encompassing signature generation and encoding, has a constant time $O(1)$. The \textsc{SliceAndEmbed} procedure iterates over $O(\mathcal{N})$ frames for slicing and embedding, resulting in a linear complexity of $O(\mathcal{N})$. Likewise, the \textsc{Verify} procedure, primarily driven by the iteration through $O(\mathcal{N})$  frames for slice reception and concatenation, also demonstrates a linear complexity of $O(\mathcal{N})$. Thus, the overall complexity is dominated by the number of frames, making it linear: $O(\mathcal{N})$.

\begin{table}[t]
    \caption{Important notations.\label{notations}}
    \centering
        \begin{tabularx}{0.42\textwidth}{l|l}
        \hline
        $ID_{AP}$ & AP's MAC address\\
        $\kappa_{\text{AP}}$ & AP's public key\\
        $\psi_{\text{AP}}$ & AP's private key\\
        $m$ & Message that the AP signs\\
        $\mathcal{S}$ & AP's digital signature\\
        $s$ & Signature size in bits\\
        $\mathcal{E}$ & Encoded digital signature\\
        $\mathcal{N}$ & Total pre\hyp{}authentication frames AP sends \\
        $P_n$ & Embedded bits on $n$th frame's preamble \\
        $f_{P}$ & Frame's preamble\\
        $f_{s}$ & Frame's preamble that contains a signature slice\\
        $\phi_{Ch}$ & Operating channel number \\
        $\phi_{Sq}$  & Sequence number of AP's last frame (EAPOL3) \\
        $t$ & The UTC time of the beacon\\
        $\mathcal{L}$ & Temporal limit\\
        \hline
        \end{tabularx}
\end{table}

\subsection{Design Details}\label{design}

We now provide the design details of our proposed scheme. 

\subsubsection{Message to Sign}
We choose the message $m = ID_{AP}\;||\;t$, where $ID_{AP}$ represents the AP's MAC address and $t$ is the UTC time that is found in the \textit{Time Advertisement} element of beacons~\cite[§\,9.4]{80211}\footnote{An AP periodically synchronizes to a UTC clock as per ITU-R Recommendation TF.460-6:2002-[B53] so that the UTC TSF offset can resolve any clock drifts~\cite[§\,11.9]{80211}.}. % This is also present in probe responses, and timing advertisement frames 
%An AP periodically synchronizes to a UTC reference clock (as per ITU-R Recommendation TF.460-6:2002---[B53]) so that the UTC TSF offset can resolve drift~\cite[Section 11.9]{80211}. 
As the signer by default sends the elements of $m$ to the verifier as part of its frames, there is no need to resend $m$.

\subsubsection{AP's Public Key}
%The station does not need the AP's public key until it receives all of the signature slices. Therefore, instead of using any extra frames, the server sends it along with the \textit{EAP} frame that contains the \gls{msk}. %We have verified that adding such a key to the \textit{EAP} frame will not exceed its maximum allowed size. Moreover, 
Irrespective of whether a station is joining a Wi-Fi network for the first time, it exchanges all of the pre\hyp{}authentication management frames explained in Section~\ref{background}. We leverage this property to eliminate the need to send the public key expiration time, as the server delivers the AP's public key to the station every time it joins the network. This eliminates the potential overhead from the delivery of the long chain of certificates and the key expiration time.

\subsubsection{Digital Signature Choices}
It is required to use a short signature since a preamble can only contain a limited number of bits. \gls*{nist} recommends maintaining a security level of at least $112$ bits \cite{nist112} which makes it challenging to find a short signature with a sufficient security level. In the EAP step, there can be as few as $13$ frames with $20$ embedded bits each; hence, the upper bound is 
%$13\times 20 = 
$260$ bits. As the AP sends the signature only, not the pair $(\mathcal{S}, m)$, the upper limit of $S$ stays at $260$ bits. We propose applying channel coding to the signature before sending it to the station to further protect it from noise, interference, and jamming, ensuring that if channel coding is applied with a rate of $r$, it should satisfy $\frac{s}{r} \leq 260$ bits total, or $\frac{s}{r\mathcal{N}}$ bits per preamble. We pick \gls{nist}-approved BLS signature~\cite{hoque2023countering}. Finally, to hash the channel number and sequence number, we chose to apply Pearson's hash variant function~\cite{pearson1990fast}.

\subsubsection{Roaming}
Our proposed scheme also supports seamless roaming under 802.11r, where the server shares the \gls{pmk} with all the APs, skipping the \gls*{eap} step and consolidating steps \circled{1} and \circled{4} in~\figurename~\ref{wifi-con-est} into only two frame exchanges. In our scheme, a target AP generates an HMAC using \gls{pmk} and sends any two random slices to the station, which then locally generates the HMAC and verifies the new AP if the two slices match any of the $13$ ones.

\subsubsection{Extra EAP Frames}
In the cases of an extra (unique) frame due to a specific EAP method, as soon as the AP needs to send any additional EAP frame it can pad or expand (e.g., using channel coding) the remaining signature bits among the subsequent frames. Once the station learns about the EAP method (and the extra frames), it can start decoding bits accordingly before verifying them.

\subsection{Protocol Features}\label{designdetails}
We now present a comprehensive overview of the security features of our proposed scheme.
\subsubsection{Temporal Limit}
We use the retransmission limit from the standard, shown in our earlier work to be $3$~\cite{hoque2022systematically}, as the authentication attempt limit $\mathcal{L}$. Brute-forcing the signature slices is then prevented by setting $\mathcal{L}=3$. The probability to successfully guess \textit{any} of the $12$ slices of a particular connection (excluding the last one) within $\mathcal{L}$ is $3.4\times 10^{-5}$. This probability is virtually zero for the last frame, EAPOL3, as the attacker would need to also successfully find a collision with the hash that is XORed with that slice (step 3). %the last preamble contains the hashed value of the final signature slice and channel number. The hash collision (with birthday paradox) probability is $9.76\times 10^{-4}$.

% Hash collision with birthday attack = 1/20^(20/2) = 1/20^10 = 9.765625e-14 

% Random guess with 
% 1st trial = 1/2^20
% 2nd trial = (1-1/2^20)(1/2^20)
% 3rd trial = (1-1/2^20-1/2^20)(1/2^20)
% success probability in 3 trials = 1st or 2nd or 3rd = 1st + 2nd + 3rd = 2.86e-6

\subsubsection{Protecting Non-cryptographically Secured Frame Elements}\label{timeboundsection}

We further protect the integrity of pre\hyp{}authentication management frames as a whole, first by distributing the slices of the same signature in sequence (chaining), and then by imposing time constraints on each frame, inspired by~\cite{brands1994distance,hussain2019insecure}, to tighten the chain. Each frame must have a cryptographically valid preamble under our scheme while the slices in these preambles are designed to be mutually dependent. As a result, it would be difficult to arbitrarily spoof one slice or modify unprotected parts of a frame with a valid preamble without being detected. Therefore, in the following, we discuss why under our time-bounding scheme one cannot easily delay or alter other frame elements by relaying a valid preamble.

When an adversary captures (and possibly blocks) a frame between a station and an AP, they may try to attach a spoofed element to the payload while keeping the original preamble (i.e., the signature slice embedded in that preamble) intact, as well as the AP's MAC address, operating channel, and sequence number~(if EAPOL3). The attacker could then send it to the station in an attempt to bypass the verification. The adversary would need to spoof/alter the target frame element(s) and ensure that the station receives the spoofed frame before the retransmission of the original one, which is a challenging task. (Note that the attacker cannot instead increment the sequence number to insert its own frame as the sequence number of the last frame should match the hashed one.). We impose time constraints on the inter-frame times to prevent a relay attack. We outline the inter-frame duration with and without a relay below.

\textit{Inter-frame duration when no relay ($t_{sta}$)--- }Let the propagation delay be $t_{prop}=d/c$~\cite{80211}, where $d$ denotes the distance between the station and an AP, and $c$ denotes the speed of light. In addition, there are other delays ($t_{other}$) involved, including transmission and processing delays, inter-frame space, and slot time. These delays (assuming transmission delay is zero) range from $0.045-20\,$ms~\cite[§\,9]{80211}. Therefore, the inter-frame time can be estimated using $t_{sta}=t_{prop}+t_{other}$. To accurately determine $t_{prop}$, the station needs the \textit{Device Location} element from a frame sent by the AP. 

\begin{figure}[h]
    \centering
    \includegraphics[scale=0.18]{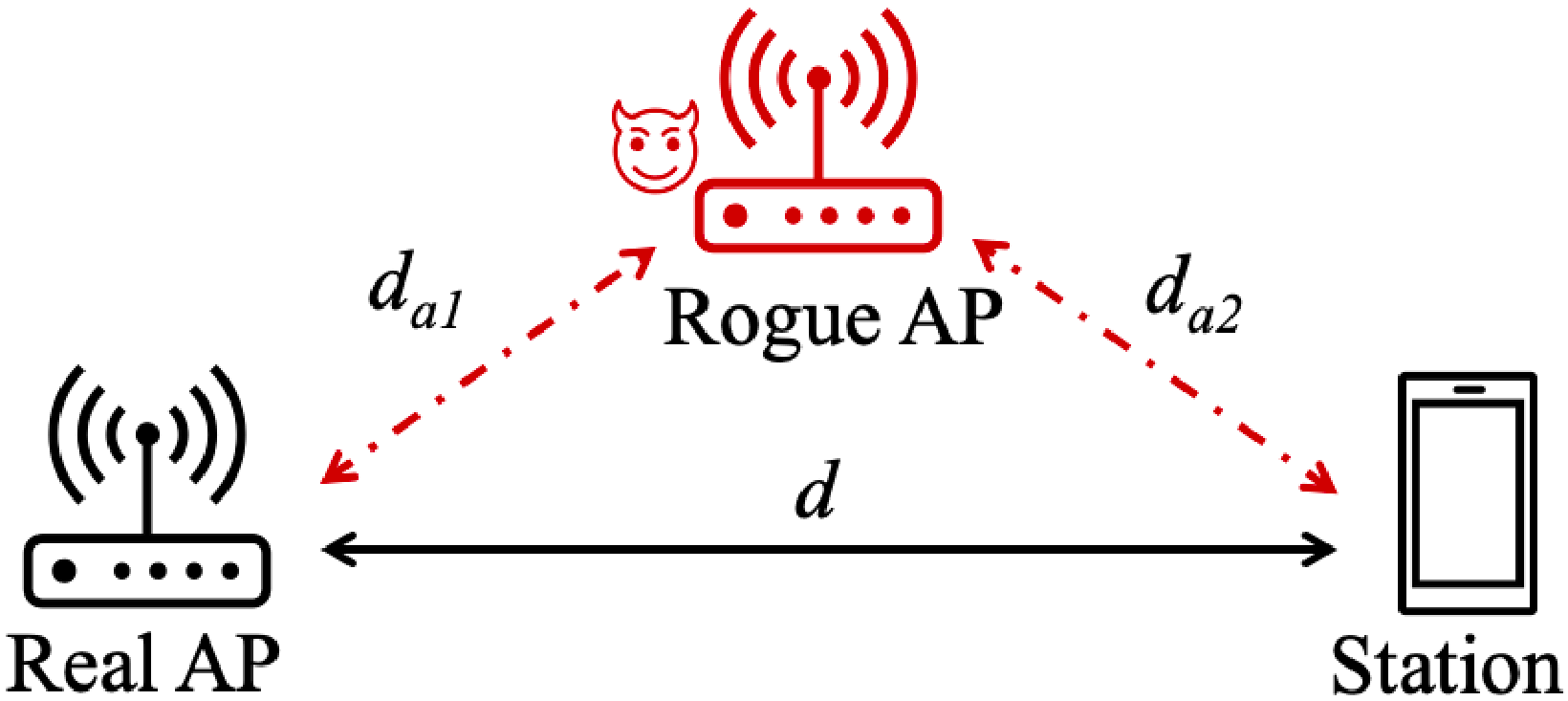}
    \caption{The adversary (relay) tries to establish a rogue AP.}\label{time-bound}
\end{figure}

\textit{Inter-frame duration when a frame is being relayed ($t_{adv}$)---} Let the distance between the real AP-adversary and adversary-station be $d_{a1}$ and $d_{a2}$, respectively (\figurename~\ref{time-bound}). A relayed frame has to travel $d_{a}=d_{a1}+d_{a2}$, and $d_{a}\geq d$. As a result, $t_{a}=c/d_{a}$. The attacker also requires some amount of processing $t_{other_a}$ ($\leq t_{other}$) and needs additional time $t_{alter}$ to perform all of its adversarial actions (i.e., block a frame, copy the slice from a preamble, attach it to a spoofed frame, and then transmit). The inter-frame time that the adversary requires, $t_{adv} = t_a+t_{other_a}+t_{alter}$, must be within the duration that the station waits before it retransmits its last frame, denoted by $t_{in}$, i.e., $t_{adv}<t_{in}$. In Section~\ref{evaluation}, we present a machine learning technique to show that the station can reliably detect if a frame is relayed even when an adversary only relays frames without altering anything (i.e., $t_{alter} = 0$). Using a learning technique, each device can determine its range of feasible $t_{in}$ values based on its computational and processing capabilities. This information can be leveraged by the device as a time constraint.

An adversary may capture a probe request from a station (and jam the \gls*{ap}) and then send this copied probe request to the AP using the original timestamp, to impersonate the station and create a \gls*{mitm} position beyond the pre-authentication phase. However, this delay in the \gls*{ce} will ultimately detected due to the tight timing margins ($t_{in}$) involved.

\subsubsection{Supporting Concurrent CEs \& Identifying Diverse Frames at the PHY layer}\label{concurrent}
When a station receives a frame during a CE, it is crucial to determine if the frame preamble contains a signature slice, indicating whether it is an expected CE frame, a beacon, an acknowledgment (ACK), or some other frame. This challenge intensifies in networks with multiple APs and stations, as the station must also discern if the received frame is from the intended AP during the CE, and whether it is indeed a CE frame or another type like a beacon, or ACK frame at the PHY layer. This needs to be achieved without referencing the MAC layer header, allowing for faster and more convenient extraction of preamble bits if the frame is a CE from the intended AP.

To address this, we use the PHY layer header’s SIG (signal) field, which contains transmission rate, frame length, etc. information. Direct observation of this information reveals no clear pattern distinguishing specific frame types or devices. Given that these are pre-authentication frames, which typically exhibit less variation compared to data frames, they often demonstrate a degree of a consistent pattern in terms of frame size, duration, and rate—details that are in the SIG field. Consequently, we propose applying \gls*{pca} to systematically analyze these nuances and effectively identify frame types and their originating devices based solely on PHY layer information. To perform \gls*{pca}, correlations between each component of the SIG field are assessed using the covariance matrix. Next, the eigenvectors of this matrix, which indicate where the most variance/information lies, are ordered from highest to lowest, allowing us to identify the most significant principal component(s). 

The SIG field contains multiple elements like transmission rate, frame length, and so on. These elements of a pre-authentication frame ($X_i$) needs to be standardized as $Z_i$, where $\mu_i$ is the mean and $\sigma_i$ is the standard deviation of each variable in $X_i$, $X_i$: $Z_i = \frac{X_i - \mu_i}{\sigma_i}$. Then, we calculate the covariance matrix $\Sigma_Z$ from the standardized data which expresses the correlation between every pair of variables in the data, where $\overline{Z}$ is the mean vector of the standardized data: $
\Sigma_Z = \frac{1}{n-1} \sum (Z_i - \overline{Z})(Z_i - \overline{Z})^T$. Next, the eigen-decomposition of the covariance matrix yields eigenvectors $\mathbf{v}$ and eigenvalues $\lambda$: $\Sigma_Z \mathbf{v} = \lambda \mathbf{v}$. Finally, sorted in descending order by eigenvalue, the top $k$ eigenvectors can selected as the principal components (PCs) based on the variance they capture: $PC_i = Z \mathbf{v}_i$, where $\mathbf{v}_i$ is the $i$th eigenvector. In Section~\ref{concurrent}, we show through our experiments that by using the PC values, a station reliably identifies frame types and the APs with 100\% accuracy, validating the effectiveness of \gls*{pca} in distinguishing diverse frame types and individual APs at the PHY layer.

\subsubsection{Tracking (and Verifying) Channel Switch(es)}\label{tracking}
%\subsection{Tracking (and Verifying) Channel Switch(es)}\label{tracking}
A station can track all valid (and invalid) channel switches. If an AP is required to change the channel after sending $P_n$ where, $n \in \{1, 2, ..., (\mathcal{N}-1)\}$, it will send the $(n+1)$th slice over the new channel:

\begin{enumerate}%[noitemsep,topsep=0pt,leftmargin=4mm]
    \item The AP sends $P'_{n+1}$ instead of $P_{n+1}$ over the new channel, where $P'_{n+1} = P_{n+1} \oplus P_{n+2} \oplus ... \oplus P_{\mathcal{N}}$. Sending the XOR of the slices that a station has yet to receive prevents an adversary from deriving and using it to validate the channel change.
    \item The rest of the frame preambles carry the regular slices: the $(n+2)$th preamble has $P_{n+2}$, the $(n+3)$th carries $P_{n+3}$, and so on.
    \item Upon receiving the last frame from the AP, and before running the verification algorithm, the station will first extract the $P_{n+1}$ by using $P_{n+2} \oplus P_{n+3} \oplus ... \oplus P_{\mathcal{N}}$.
    \item The station recovers the operating channel number $\phi_{Ch}$ and last frame's sequence number $\phi_{Sq}$ from $P_{\mathcal{N}}$ and combine all of the signature slices.
    \item If the channel change occurs after $P_{\mathcal{N}-1}$, the AP will simply hash $P_{\mathcal{N}}$ with the PTK (as AP has the PTK by then). The station will first recover $P_{\mathcal{N}}$ with the PTK, then will reconstruct the full signature, and finally will verify it.
\end{enumerate} 

Although a cryptographically hashed channel number verification mechanism confirms that both the AP and the station are indeed on the same channel at the time of the transmission of the last pre\hyp{}authentication frame (EAPOL3) by the AP, it cannot verify if there have been any channel switch(es) prior to that point, including when an adversary forces invalid channel switch(es), then returns to the original one before the AP’s EAPOL3 transmission. The mechanism of XORing unseen slices can detect all such valid and malicious channel switch attempts. The station tracks valid/invalid channel switches during setup and reports to the network administrator once a protected data session begins. An invalid switch signals the presence of an adversary, while multiple valid switches suggest unstable connections in specific areas of a public place, such as a corner in an airport.

\textit{CSA Element Reception Acknowledgment. }
If a station receives a \gls{csa} before sending its $n$th management frame, it uses the \gls{epmod} technique to include the next frame's channel ($P_{n+1}$) in the current preamble ($P_{n}$). This prevents the AP and station from diverging channels during setup, similar to the Query exchange in~\cite{vanhoef2018operating}. We note that any MitM (relay) attack resulting from spoofing this acknowledgment will be detected by the station. %This is similar to the OCV technique's SA Query exchange~\cite{vanhoef2018operating}, except our proposed solution does not require sending any extra frames for this purpose. All stations will perform this if they receive any beacon with CSA.

\subsubsection{Frame-by-Frame Preamble Authentication}As discussed earlier, both symmetric and asymmetric solutions can protect the chain of pre\hyp{}authentication frames. However, only a symmetric key-based approach can immediately detect preamble spoofing attacks or a corrupted preamble upon receiving \textit{each} preamble, as both parties will have access to the symmetric key shared during the initial EAP process to generate the same HMAC. 
In contrast, under the asymmetric key-based approach, the station does not have access to the AP's private key to generate identical signature slices. %The station can generate the same HMAC using a symmetric key locally, which the server shares during the initial EAP. If an adversary injects a spoofed preamble or somehow a preamble becomes corrupted, the station can immediately identify a discrepancy between its local copy and the version in the potentially spoofed preamble, leading it to reject that preamble. 
When a station joins a network for the first time, however, it will not be able to use a symmetric-key approach since the symmetric key will be available only after the first EAP process.

% \subsubsection{Frame-by-Frame Preamble Authentication}As discussed earlier, both symmetric and asymmetric solutions can protect pre\hyp{}authentication frames. However, to detect preamble spoofing attacks frame by frame (i.e., immediate slice verification), an asymmetric key solution is not appropriate since a station does not have AP's private key to generate the same signature. A station can verify HMAC slices as frames arrive since both parties have access to the symmetric key. The station can generate the HMAC locally and compare each slice upon receiving \textit{each} preamble to verify its integrity. When a station joins a network for the first time, however, it will not be able to use this approach since the symmetric key will be received during the EAP process.

%\subsubsection{Attempts to Extend MitM Control beyond Connection Establishment Phase}An adversary may copy a probe request from a station (and jam it) and then delay its connection setup with the AP to maintain their \gls*{mitm} position beyond the pre-authentication phase. During this disruption, the attacker sends this copied probe request to the AP using the original timestamp, in an attempt to impersonate the station. However, this attack ultimately fails during the EAP authentication, as the attacker does not possess the station's credentials.

\subsection{Limitations of Alternative Approaches}\label{limitation}
\textit{One digital signature for each pre\hyp{}authentication frame---} If an AP generates one signature per frame, it will be costly in terms of the signature generation time (at the AP’s end) and verification time (at the station’s end), where the latter cannot even be performed before the station receives the AP's (public) key. Also, it will require increasing the size of every frame, which will add delay and communication overhead to the joining process and further conflicts with the IEEE 802.11ai goal for fast link setup (FILS)~\cite{80211ai}.

\textit{One digital signature of a message digest and sending it over the last frame sent by the AP---} We have also considered a TLS-like solution to verify an AP's legitimacy. In TLS, application layer data across all the packets is converted into a single message digest using \gls{hmac} with a symmetric key and sent to the receiving end to protect the payload integrity during the entire process. %For confidentiality, the digest is encrypted using the private key of an asymmetric key pair (forming a digital signature). The signature then is decrypted by 
The receiver compares it with a locally generated digest. A similar solution for our problem can be as follows. The ``message'' here is the payload of all pre\hyp{}authentication frames sent by the AP combined. After creating the digest using a symmetric key, the AP will send it in its last pre\hyp{}authentication (EAPOL3 message) frame payload. The station can verify it by comparing it against a locally generated digest. This method will not only require an extension of one frame but more importantly, cannot protect against preamble-based spoofing or multi-channel MitM attacks at the PHY and MAC layers, as discussed in Sections~\ref{introduction} and~\ref{background}. Our aim is not only to protect the connection establishment from a multi-channel MitM, but also from the PHY-layer attacks such as starvation attack~\cite{gvozdenovic2020truncate}, frame-detection attack, and data alteration attack~\cite{zhang2021preamble} that a TLS-like solution (merely a message digest) cannot protect. These attacks happen at the PHY or MAC layer where the upper layers are blind. Only a PHY-layer approach can prevent such attacks.

%% file: 5.evaluation.tex
\section{Experimental \& Formal Security Evaluation}\label{evaluation}

In this section, we show that our proposed method excels in communication performance and security, tested under current Wi-Fi standards and through formal security analysis.

\subsection{Experimental Evaluation}
In the following, we systematically detail the experimental evaluation of our proposed technique.

\subsubsection{End-to-end System Performance Evaluation} We now provide the implementation details and end-to-end communication performance evaluation of our proposed technique. 

\noindent \textit{Metrics--} To quantify the end-to-end performance of our proposed work, we use the following metrics: signature/HMAC generation/verification times, signature \gls{sr} (defined below), \gls{ber}, extraction time, and the total connection establishment time with our solution. The signature \gls*{sr} is the ratio of the number of correctly received signatures once reassembled to the total signatures. It does not indicate how many bits are received correctly in one signature. Whether only one or all the bits are received erroneously, it will be counted as \textit{unsuccessful}. The extraction time is the time to recover one signature slice from a preamble. The total connection establishment time includes bit extraction,  generation/verification times, and the existing Wi-Fi system's connection establishment time.

\noindent \textit{Implementation to Support IEEE 802.11ax--}
Commercial APs do not allow modifying their firmware, where the preamble is implemented, and that limits our ability to use a testbed to fully evaluate the proposed scheme. Open-source full-stack Wi-Fi implementations, such as Openwifi~\cite{openwifigithub}, also do not support the 802.11ac/ax, and their FPGA implementations of the preamble are not supported on easily accessible boards either. That said, the preamble part of the \textit{gr-ieee802-11} library, an open-source GNU Radio module designed to support IEEE 802.11a, is modifiable; however, it lacks support for 802.11ac/ax. To address this, we modified it in the following way to not only support IEEE 802.11ax but also to implement a preamble bit extraction technique at the receiver.

\fbox{1} Since the library only supports 802.11a at 20 MHz, we expanded it to also support 802.11ax over 20 MHz of bandwidth based on the standard requirement~\cite{80211}. For example, the IEEE 802.11a standard uses an FFT size of 64 which is utilized for Orthogonal Frequency-Division Multiplexing (OFDM) in the physical layer to modulate data over multiple subcarrier frequencies. In contrast, IEEE 802.11ax significantly increases the FFT size to accommodate higher data rates and increased efficiency. For 20 MHz channels, it uses a 256-point FFT. The larger FFT size in 802.11ax allows for finer granularity in frequency division, which is essential for its Orthogonal Frequency-Division Multiple Access (OFDMA) technologies and improves subcarrier spacing. 

\fbox{2} The original library discards the preamble after detecting a new frame. Since our proposed technique utilizes the preambles, we added that functionality to the library. The received preamble then goes through the Fourier transform and channel estimation process like the rest of the frame. 

\fbox{3} Then, we implemented the \textit{\gls{epmod}} technique as described in~\cite{zhang2021adaptive} to extract the embedded bits from the received 802.11ax frames. As the firmware of a commercial AP cannot be modified, we assume for simplicity that the embedded signature bits are constant. Our experimental setup, involving commercial APs, laptops, and USRP X310, advances from our initial work only on IEEE 802.11ac with a USRP B210.

\begin{figure}[t]
\centering
    \begin{tikzpicture}
    \begin{axis}[
        xlabel={Number of CE Frames},
        ylabel={SR},
        xmin=12.8, xmax=15.2,
        ymin=0.3, ymax=1,
        xtick=data,
        axis y line*=left,
        ylabel near ticks, yticklabel pos=left,
        grid=major,
        legend pos=south west,
        width=6.5cm,
        height=3.8cm
    ]

    % Plotting Success Rate values
    \addplot [color=blue, mark=o, line width=1.0pt] coordinates {
        (13, 0.9) (14, 0.91) (15, 0.915)
    };
    \addlegendentry{SR}

    \end{axis}
    
    \begin{axis}[
        axis y line*=right,
        axis x line=none,
        ylabel={BER},
        ymin=0.0006, ymax=0.01,
        ymode=log,
        ylabel near ticks, yticklabel pos=right,
        legend pos=south east,
        width=6.5cm,
        height=3.8cm
    ]

    % Plotting BER values
    \addplot [color=red, mark=square, line width=1.0pt] coordinates {
        (13, 0.0025) (14, 0.0022) (15, 0.0016)
    };
    \addlegendentry{BER}
    
    \end{axis}
    \end{tikzpicture}
\caption{SR and BER across different numbers of CE Frames.}
\label{fig:SR_and_BER}
\end{figure}
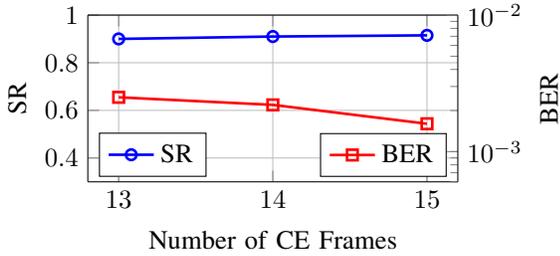

\begin{table}[t] 
\caption{Number of frames needed to send $160$ bits of signature ($N_{frames}$), number of bits/slice ($N_{bits/slice}$), the total bit extraction time ($T_{extract}$), and total connection establishment time ($T_{CE}$).}\label{resulttable}
\centering
\begin{tabularx}{0.49\textwidth}{ 
| >{\centering\arraybackslash}X 
| >{\centering\arraybackslash}X 
| >{\centering\arraybackslash}X 
| >{\centering\arraybackslash}X | }
    \hline
    $N_{frames}$& $N_{bits/slice}$ & $T_{extract}$ (ms) & $T_{CE}$ (ms)\\ 
    \hline
    $13$& $13$ & $0.26$ & $306.54$\\
    \hline
    $14$ & $12$ & $0.28$ & $306.56$\\
    \hline
    $15$ & $11$ & $0.30$ & $306.58$\\
    \hline
\end{tabularx} 
\end{table}

\noindent \textit{Digital Signature--} We measure the signature generation and verification times using the Pairing-Based Cryptography (PBC) library~\cite{pbc}. The BLS algorithm takes $0.65\,ms$ (standard deviation (stddev) $0.07$\,ms) and $5.63\,ms$ (stddev $0.33$\,ms) to generate and verify a signature, respectively.

\noindent \textit{Bit extraction time \& total CE time--} 
To assess the time required for connection establishment in the existing IEEE 802.11ax Wi-Fi networks, we analyze the CE phase utilizing Wireshark to log the timing of frame exchanges. Our findings indicate that the average duration for CE is approximately $300$ ms (stddev $14.3$ ms). On average the time taken for a frame to be transmitted from an AP to a station (or the reverse) is observed to be $9.73$ ms, with the average time between successive frames (inter-frame time, $t_{in}$) originating from an AP being $18.66$ ms.

Our experimental data reveal that the time taken to process and extract data from each frame by the receiving station is substantially less than the time gap between consecutive frames (about $18.66$ ms on average), suggesting that bit extraction from preamble occurs promptly after their reception, without incurring additional delays. According to Table~\ref{resulttable}, even in the least efficient scenario—where a station begins to extract the preamble bits only after all pre-authentication frames have been received—the total average extraction time is minimal- merely $2.19\%$ of the average connection time of 300 ms of the existing CE in Wi-Fi.

\noindent \textit{Success rate \& Bit error rate--} Our result shows (see~\figurename~\ref{fig:SR_and_BER}) that the \gls*{sr} is around $92\%$. Considering that each 802.11ax preamble can embed up to $20$ bits, there are still $7$ bits/frame remaining that we can use for error-correction and improve \gls*{sr} and reduce \gls*{ber}. While error-correcting methods could be utilized to improve the \gls*{sr}, we could not apply them due to the restriction of the commercial AP-- which prevented us from altering any preamble bits. Therefore, via simulation, we show in Section~\ref{error_coding} that incorporating error correction can significantly enhance the \gls*{sr}. %No coding scheme is applied for our testbed experiments because modifying a commercial AP's firmware is not possible. However, we showed via simulations in Section~\ref{error_coding} that channel coding improves the \gls*{ber} and \gls*{sr}. 

\subsubsection{Relay Attack Demonstration \& Detection}

In this experiment, a relay attack presents an attacker positioning themselves between an AP and a station, capturing and forwarding frames without any alteration. This attack is particularly insidious because the attacker does not modify the contents of the frames; instead, they simply relay them from the AP to the station. The primary aim of this evaluation is to test the effectiveness of our time-bound detection technique. By demonstrating that our system can identify frames that are merely forwarded—and thus, have not been tampered with in transit—we posit that it can more reliably detect any relayed frame, including those in which an attacker may introduce changes or spoof any frame element.

\noindent \textit{Metrics--} To evaluate the efficacy of our time-bound relay attack detection mechanism, we considered the following metrics: accuracy, f1-score, true positive/negative rate (TPR/TNR), and positive/negative predictive value (PPV/NPV). Accuracy reflects the overall correctness in detecting both benign and relayed frames. The f1-score provides a nuanced view, balancing the system's precision/PPV (its success in correctly identifying relayed frames) and recall/TPR (its ability to detect all relayed frames), to assess the trade-offs between false positives and false negatives. NPV assesses the system's effectiveness in verifying frames as benign.

\begin{figure}[t]
\centering
    \subfloat[Relay attack and detection experiment setup: attack devices inside the room, AP outside, testing across three station locations (D1, D2, D3). The station is positioned at D2 on the cart in this picture.]{\includegraphics[scale=0.22]{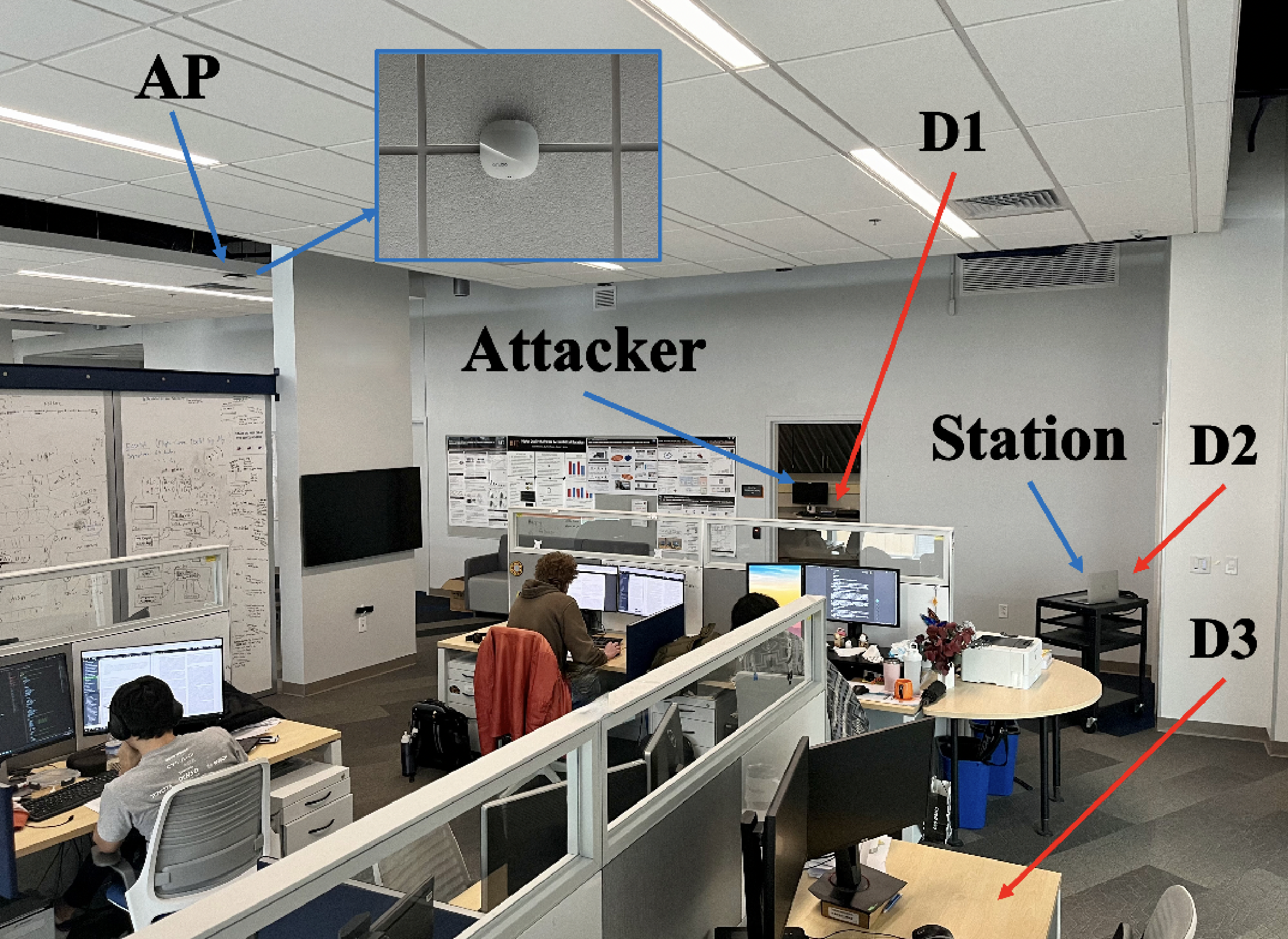}\label{exp_relay_setup}}\\
    \subfloat[USRP X310 connected with the host machine and PCIe as the relay setup for the attacker.]{\includegraphics[scale=0.11]{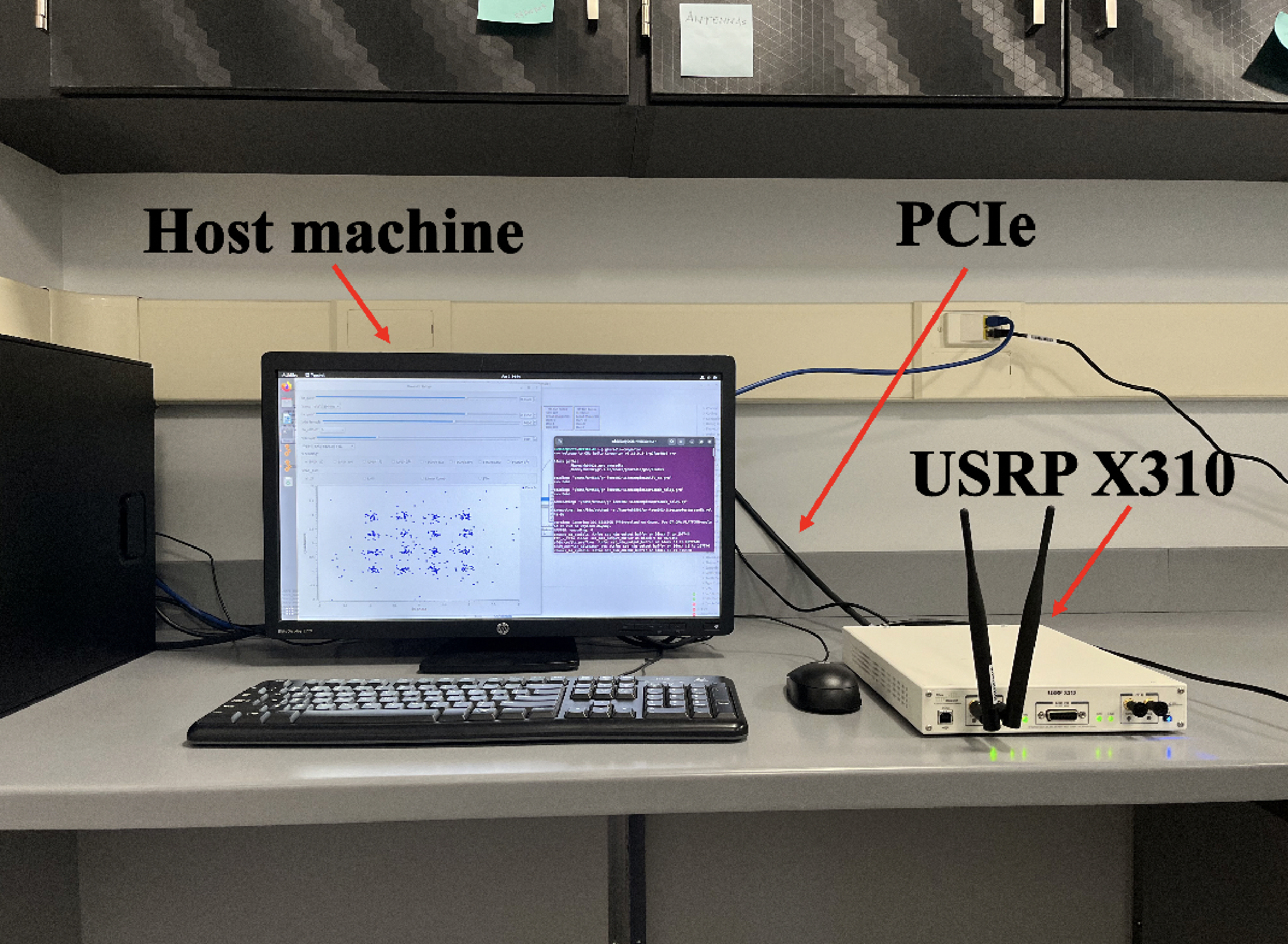}\label{exp_relay_device}}\;
    \subfloat[Inter-frame times of a set number of benign versus relayed frames across D1, D2, and D3.]{\includegraphics[scale=0.175]{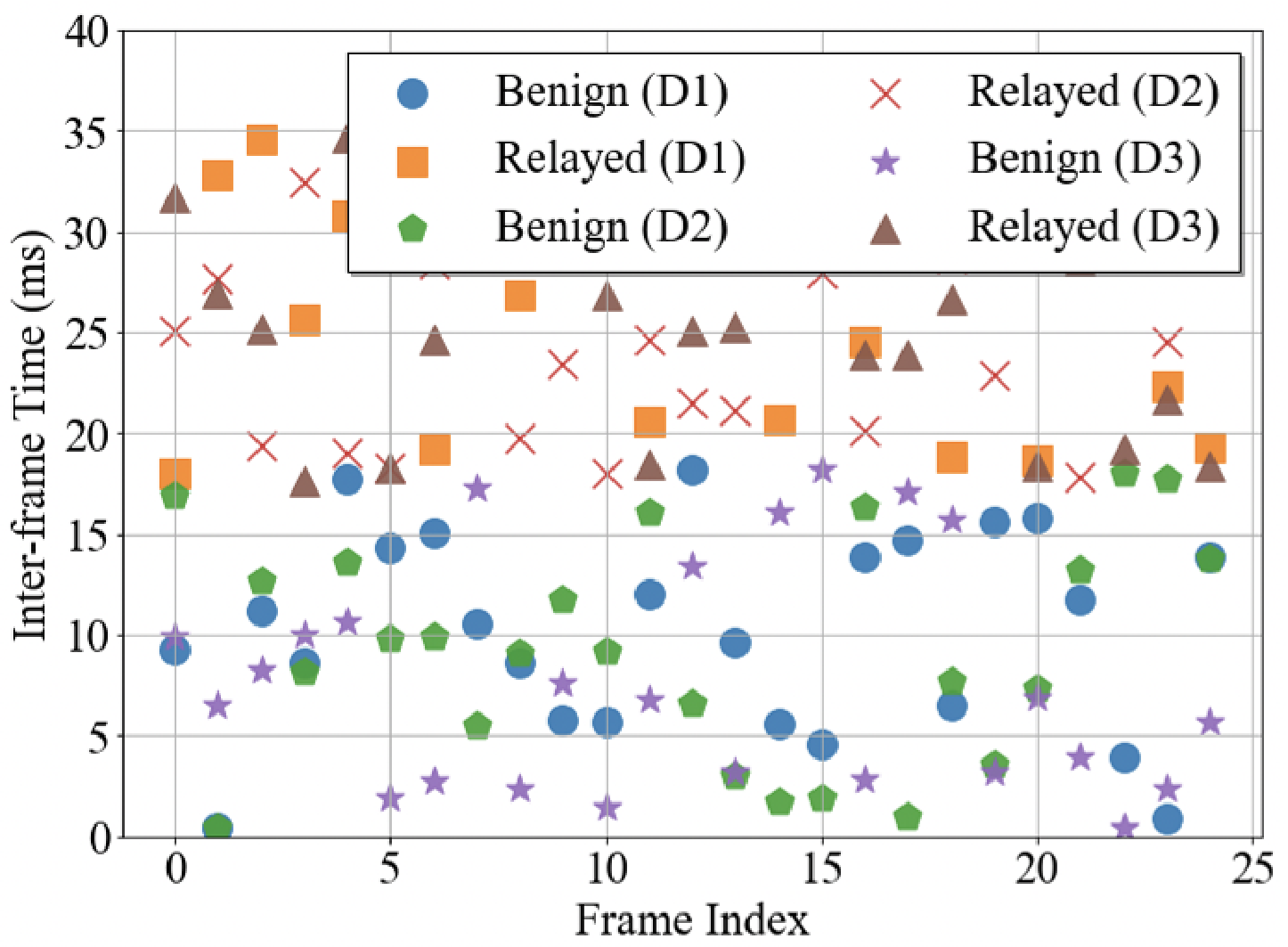}\label{deltadata}}
    \caption{Relay attack and detection experiment.}\label{relayexp}
\end{figure}

\begin{table}[t]
\centering
\caption{Relay attack detection performance.}
\label{relayresults}
\begin{tabular}{@{}|c|c|c|c|c|c|c|@{}}
\hline
Setup & Accuracy & F1-score & TPR & TNR & PPV & NPV \\
\hline
D1 & 0.975 & 0.974 & 1.0 & 0.952 & 0.950 & 1.0 \\
\hline
D2 & 0.975 & 0.978 & 0.957 & 1.0 & 1.0 & 0.944 \\
\hline
D3 & 0.975 & 0.979 & 1.0 & 0.938 & 0.960 & 1.0 \\
\hline
\end{tabular}
\end{table}

\noindent \textit{Devices Used--} We utilized a high-performance commercial Aruba AP supporting an IEEE 802.11ax enterprise Wi-Fi environment. The station device was a MacBook Air (M1, 2020) with 8 GB RAM. The relay attacks were orchestrated using an Ubuntu-based relay host machine with 32 GB RAM and an Intel Core i7 6700 processor at 3.4 GHz across 8 cores, equipped with an Ettus Research USRP X310 connected via PCIe to facilitate rapid signal processing (see \figurename~\ref{relayexp}\subref{exp_relay_device}). 

\noindent \textit{Implementation \& Experimental Setup--} To accurately demonstrate such a relay attack, it was essential to develop a relay mechanism capable of processing and forwarding received frames with minimal delay. This ensures that the relayed frame reaches the intended station within a critical time frame ($t_{in}$), bypassing the time-bound detection threshold designed to flag malicious activity. There, developing a high-speed transceiver was central to this endeavor. 

To do that, we designed a transceiver in GNU Radio that contains a receiver block to capture Wi-Fi frames, apply noise measurement and channel estimation, demodulate, and then directly send the I/Q samples to the transmitter block. Understanding the time-bound, or $t_{in}$, between an AP and a station is crucial for both attackers aiming to remain undetected and defenders seeking to identify such breaches. So, the transmitter modulates and transmits the samples to the victim station. To confirm that our replay attack is working correctly, we did not block the original frame, rather we checked at the receiving end if it received a specific frame is received two (one original and one relayed) at the victim station. 

As we discussed in Section~\ref{timeboundsection}, an adversary needs $t_{adv}=t_a+t_{other_a}+t_{alter}$ amount of time if it wants to alter or spoof any frame element while relaying. Now we consider that it only relays the frames-- it does not alter anything, then $t_{alter} = 0$. It is the worst-case scenario for our detection technique. We consider both scenarios with and without the presence of an adversary. Our goal is to collect $t_{adv}$ and $t_{sta}$ and use the collected inter-frame duration to assess whether it shows any distinction between relayed and non-relayed frames.

\noindent \textit{Analyses \& Results--}  We conducted our experiment across three different station locations as shown in Figure~\ref{relayexp}\subref{exp_relay_setup}. As shown in \figurename~\ref{relayexp}\subref{deltadata}, our collected values indicate that the presence of an adversary involves additional latency ($t_{other}$). The visualization of the inter-frame time differences of each sample with and without adversarial attempts indicates that a station can identify if a frame is relayed by looking at the inter-frame duration, as an adversary will always require an additional $t_{other}$ even if it is only relaying the frame. 

To further validate our observation that a station can identify a relay attack using the inter-frame duration, we apply machine learning techniques to our collected dataset of inter-frame duration with ($t_{adv}$) and without ($t_{sta}$) relay attack. We split this dataset into train-test parts and use \textit{Scikit-learn} Python library to implement and evaluate random forest (RF). Our results achieve $97.35\%$ accuracy and $98\%$ f1-score across all of the three locations (D1-D3) as listed in Table~\ref{relayresults} with other high-performance scores.

\subsubsection{Performance under Concurrent CEs} 

We distinguish frame types at the PHY layer by analyzing the SIG field in the PHY header, avoiding MAC layer inspection. This approach leverages the consistent pattern in size, duration, and rate of pre-authentication frames, detailed in the SIG field, enabling a station to identify frame types like beacons, acknowledgments, or \gls*{ce} frames solely through PHY layer information. In the following, we describe our experiment and result details.

To do that, we captured and analyzed Wi-Fi frames, focusing specifically on pre-authentication frames to isolate essential information from the SIG field of the PHY layer header. Employing \gls*{pca}—a statistical method that identifies directions of maximum data variance—we show, in the following that a station can differentiate among multiple APs and distinct frame types.

\noindent \textit{Experimental Setup \& Devices Used--} For our experimental setup, we considered five APs representing different SSIDs, labeled A to E. Table~\ref{device_list} lists the stations we considered for this experiment. We conducted our experiment in the same room as shown in Figure~\ref{relayexp}\subref{exp_relay_setup}, where the stations received frames from different SSIDs. To ensure multiple concurrent \glspl{ce}, each station began connecting to an AP simultaneously. We repeated these attempts with each of the five APs one at a time. We captured the pre-authentication frames exchanged between each AP-station pair. 

\begin{figure}[t]
    %\vspace{-0.2in}
    \subfloat[PC values of APs (SSIDs) A to E.]{\includegraphics[scale=0.225]{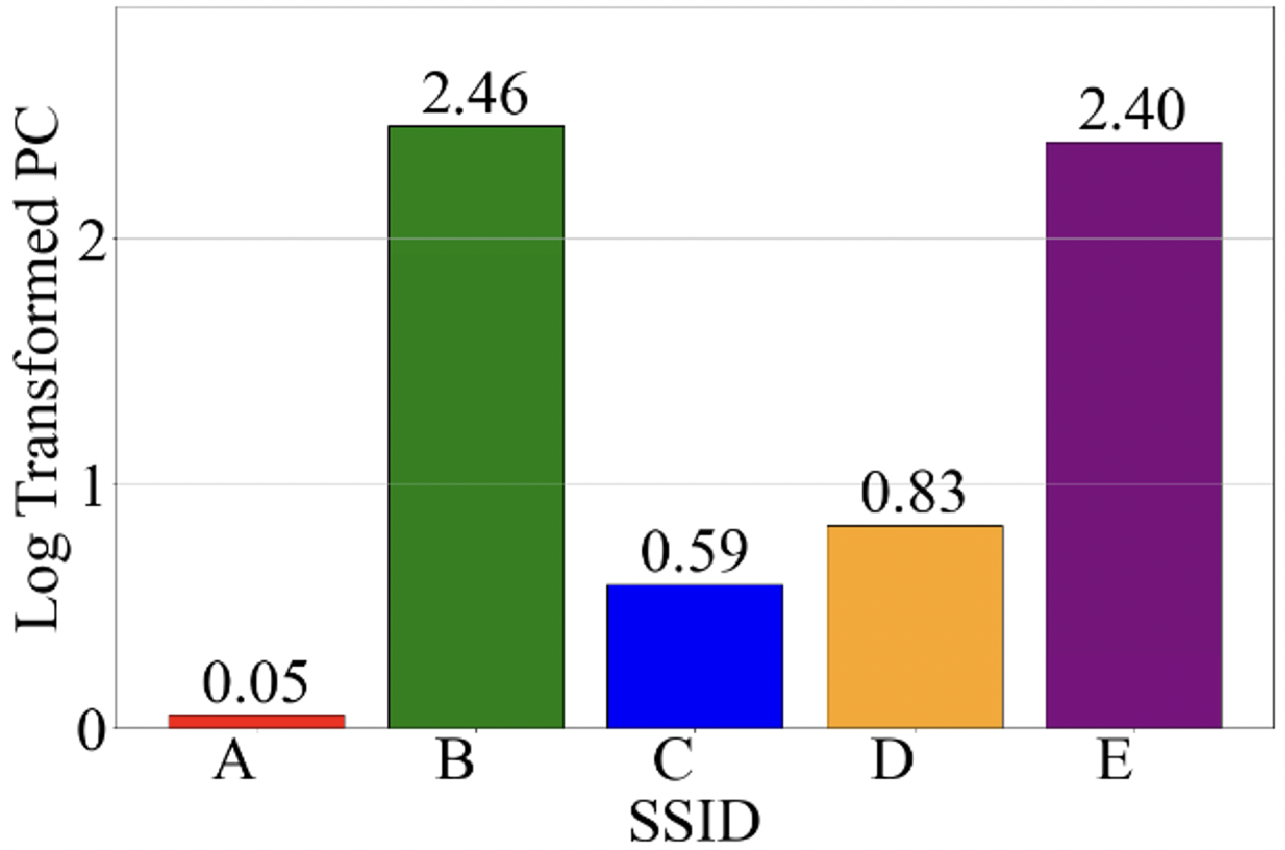}\label{pc_ap}}\;
    \subfloat[Confusion matrix (CM).]{\includegraphics[scale=0.24]{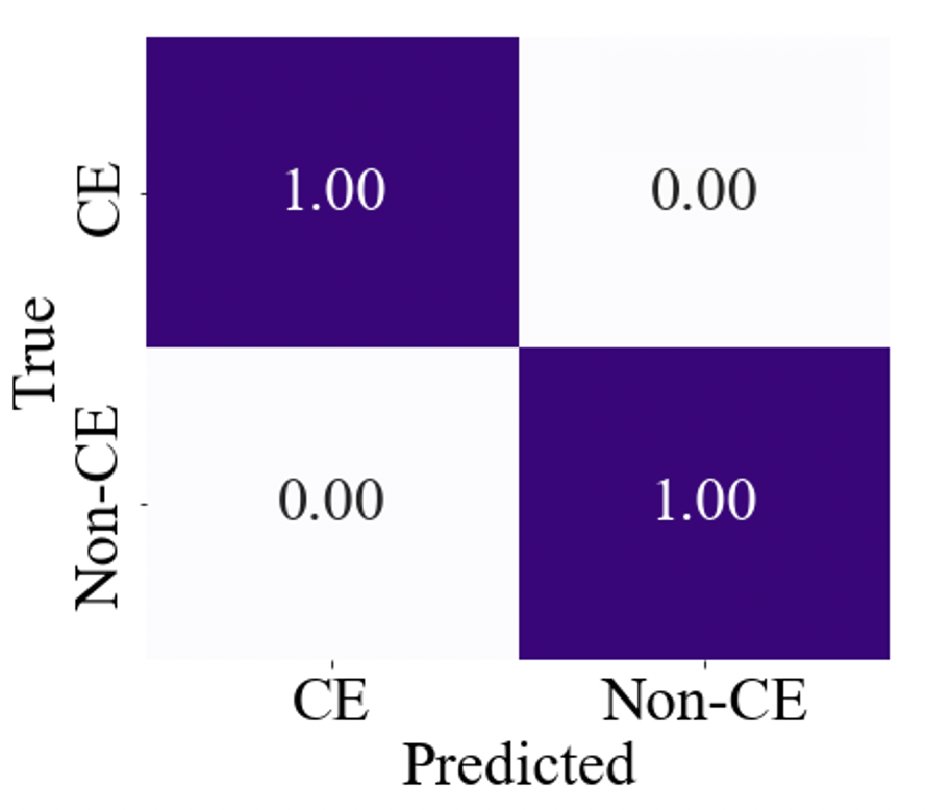}\label{cm_ce}}
    \caption{PC values for APs demonstrate that stations can accurately distinguish them, with the confusion matrix showing its ability to correctly identify CE and non-CE frames.}\label{concurrent_ce}
   % \vspace{-0.2in}
\end{figure}

\noindent \textit{Analyses \& Results--} As already mentioned, we use \gls*{pca} to analyze the SIG field from the PHY layer header of the captured frames. We show in ~\figurename\ref{concurrent_ce}\subref{pc_ap} that the average PC values for each of the five APs are unique, allowing stations to differentiate between the APs. We also utilized the PC values for frames and demonstrated that the station can identify with 100\% accuracy (see ~\figurename\ref{concurrent_ce}\subref{cm_ce}) whether a received frame is a CE frame or a non-CE one.

\begin{table}[t]
\caption{Devices used for our analysis of Wi-Fi's CE phase.}
\centering
\begin{tabular}{|l|l|l|}
\hline
\textbf{Device Type} & \textbf{Model} & \textbf{Operating System} \\ \hline
Laptop & MacBook Pro (M1, 2021) & macOS Monterey \\ \hline
Laptop & MacBook Air (M1, 2020) & macOS Big Sur \\ \hline
Laptop & Dell XPS 15 & Windows 10 \\ \hline
Laptop & Lenovo ThinkPad X1 Carbon & Windows 10 \\ \hline
Smartphone & iPhone 14 & iOS 15 \\ \hline
Smartphone & iPhone 13 & iOS 15 \\ \hline
Smartphone & iPhone 11 Pro & iOS 14 \\ \hline
Smartphone & Samsung Galaxy S21 & Android 11 \\ \hline
\end{tabular}
\label{device_list}
\end{table}

\begin{figure}[t]
\centering
    \begin{tikzpicture}
    \begin{semilogyaxis}[
        xlabel={SNR (dB)},
        ylabel={BER},
        xmin=0, xmax=15,
        ymin=1e-5, ymax=1,
        xtick={0,3,...,16},
        ymode=log,
        grid=major, 
        legend pos=north east,
        legend style={at={(0.5,-0.7)},anchor=north,legend columns=2}, 
        width=5.8cm,
        height=3.1cm
    ]

    % Existing AWGN BER data
    \addplot [color=orange, dashed, mark=*, line width=1pt, dash pattern=on 10pt off 5pt] 
        coordinates {
        (0, 0.1586) (3, 0.0791) (6, 0.0232) (9, 0.00024) (12, 0) (15, 0)
        };
    \addlegendentry{AWGN}

    % New Rayleigh BER data
    \addplot [color=cyan, dashed, mark=square, line width=1pt, dash pattern=on 10pt off 5pt] 
        coordinates {
        (0, 0.2852) (3, 0.2120) (6, 0.1295) (9, 0.0557) (12, 0.0123) (15, 0.0007)
        };
    \addlegendentry{Rayleigh}
    
    \addplot [color=purple, mark=x, line width=1pt] 
        coordinates {
        (0, 0.3648) (3, 0.0377) (6, 4e-5) (9, 0) (12, 0) (15, 0)
        };
    \addlegendentry{AWGN (coded)}

    \addplot [color=blue, mark=triangle, line width=1pt] 
        coordinates {
        (0, 0.4942) (3, 0.4662) (6, 0.2433) (9, 0.0074) (12, 0.0001) (15, 0.0000)
        };
    \addlegendentry{Rayleigh (coded)}

\end{semilogyaxis}
\end{tikzpicture}
\caption{Impact of \gls{ber} under different SNR levels.}
\label{ber_coded}
\end{figure}
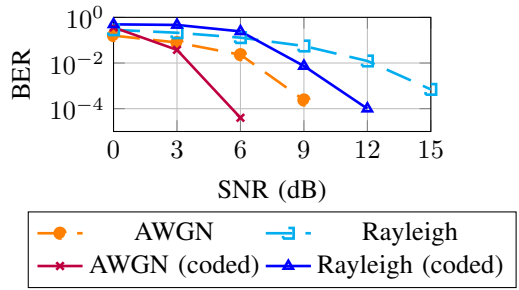

%\vspace{-0.3in}

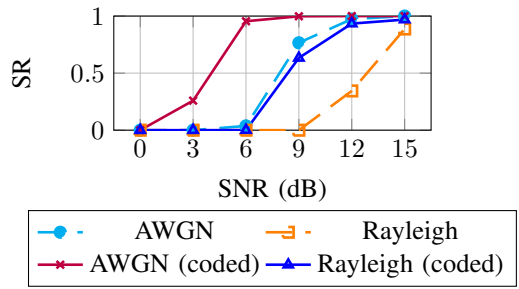
\begin{figure}[t]
\centering
    \begin{tikzpicture}
    \begin{axis}[
        xlabel={SNR (dB)},
        ylabel={SR},
        ymin=0, ymax=1,
        ytick={0,0.5,...,1},
        xtick={0,3,...,16},
        grid=both,
        legend pos=north east,
        legend style={at={(0.5,-0.7)},anchor=north,legend columns=2}, 
        width=5.8cm,
        height=3.1cm
    ]

    % Existing AWGN SR data
    \addplot [color=cyan, dashed, mark=*, line width=1pt, dash pattern=on 10pt off 5pt] 
        coordinates {
        (0, 0) (3, 0) (6, 0.0374) (9, 0.7658) (12, 0.9736) (15, 1)
        };
    \addlegendentry{AWGN}

    % New Rayleigh SR data
    \addplot [color=orange, dashed, mark=square, line width=1pt, dash pattern=on 10pt off 5pt] 
        coordinates {
        (0, 0) (3, 0) (6, 0) (9, 0) (12, 0.3440) (15, 0.8867)
        };
    \addlegendentry{Rayleigh}

    \addplot [color=purple, mark=x, line width=1pt] 
        coordinates {
        (0, 0) (3, 0.2591) (6, 0.9547) (9, 0.998) (12, 1) (15, 1)
        };
    \addlegendentry{AWGN (coded)}
    
    \addplot [color=blue, mark=triangle, line width=1pt] 
        coordinates {
        (0, 0) (3, 0) (6, 0.0008) (9, 0.6333) (12, 0.934) (15, 0.9691)
        };
    \addlegendentry{Rayleigh (coded)}

\end{axis}
\end{tikzpicture}
\caption{Impact of SR under different SNR levels.}
\label{sr_coded}
\end{figure}

\renewcommand{\arraystretch}{1.2}
\begin{table*}[h]
\centering
\caption{Symbolic model checking: finite state machine (FSM) of a station and its transitions.}
\label{formalmodel}
\begin{tabularx}{\textwidth}{|>{\hsize=.1\hsize}X|>{\hsize=.2\hsize}X|>{\hsize=1.7\hsize}X|c|}
\hline
${\tau}$& Action & Condition &\\  \cline{1-3}
${\tau_1}$ & $\alpha_1 \;\text{to}\; \alpha_2$& ${\mathcal{C}^{as1}}$  &  \multirow{6}{*}[1.65em]{\includegraphics[scale=0.16]{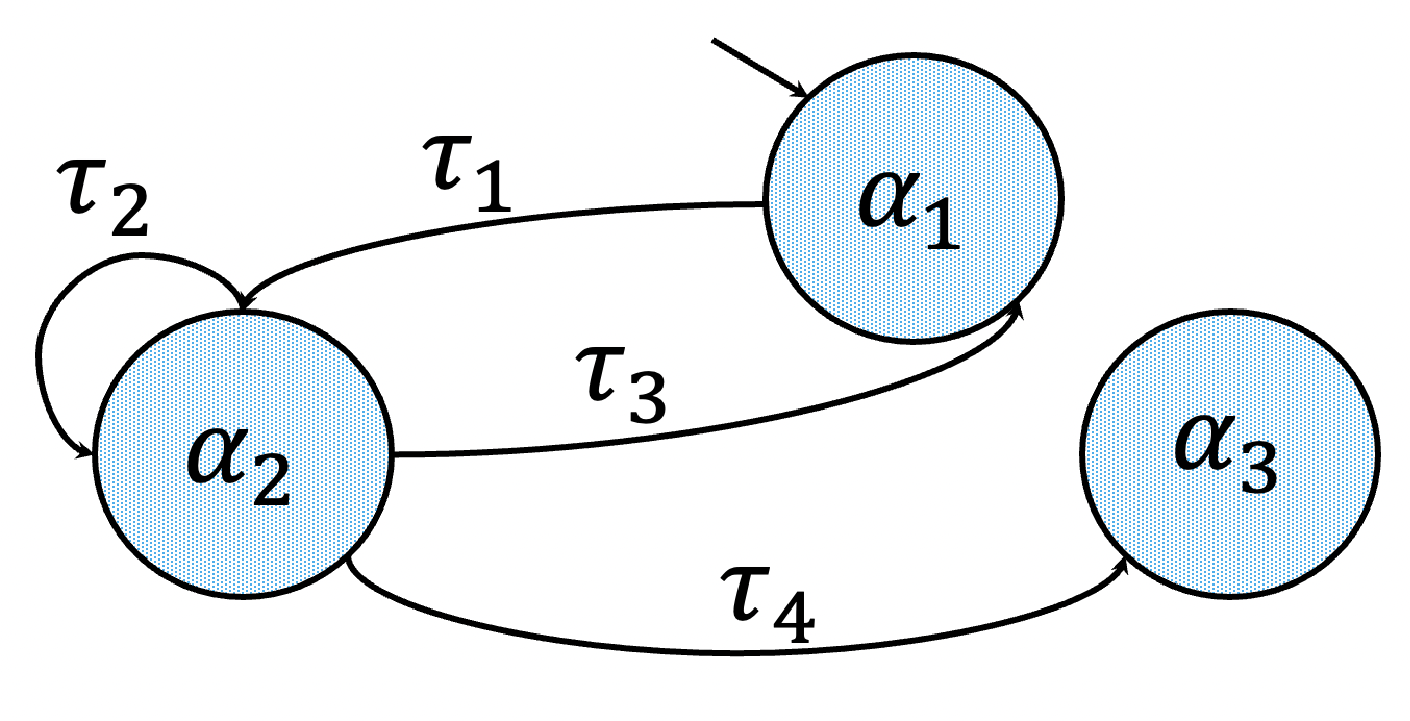}}\\ \cline{1-3}
${\tau_2}$ & $\alpha_2 \;\text{to}\; \alpha_2$ & $(\neg \mathcal{C}^{asN} \land \mathcal{L} \land (\neg \phi_{Ch} | \neg \phi_{Sq} | \neg f_s)) | (\neg \mathcal{C}^{saN}  \land t_{\text{in}} \land \mathcal{L} \land (\neg \phi_{Ch} | \neg \phi_{Sq} | \neg f_s))$ &  \\ \cline{1-3}
${\tau_3}$ & $\alpha_2 \;\text{to}\; \alpha_1$& $(\neg{\mathcal{C}^{saN}} \land \neg{t_{in}} \land \neg\mathcal{L} \land (\neg\phi_{Ch}| \neg\phi_{Sq} | \neg{f_s})) | (\neg{\mathcal{C}^{asN}} \land \neg\mathcal{L} \land (\neg\phi_{Ch}| \neg\phi_{Sq} | \neg{f_s}))$ &  \\ \cline{1-3}
${\tau_4}$ & $\alpha_2 \;\text{to}\; \alpha_3$ &  $\mathcal{C}^{saN} \land \mathcal{L} \land \phi_{Ch}\land \phi_{Sq} \land f_s $ &  \\ \hline
\end{tabularx}
\end{table*}

\renewcommand{\arraystretch}{1}
\begin{table}[h]
\caption{Formal Security Verification}
\label{formalverification}
\centering
\begin{tabular}{|p{0.1\textwidth}|p{0.33\textwidth}|}
\hline
\rowcolor{gray!30} 
\multicolumn{2}{|c|}{\textbf{\circled{A} Model checker (MC) -- Attacks checked} }\\
\hline
\rowcolor{gray!20} 
\textbf{Attack} & \textbf{Adversary actions} \\
\hline
Spoofing & Adversary sends a fake CSA announcing channel switch to $y$ to STA FSM (CSA-based MitM). \\
\hline
Relay & Forces AP FSM to switch channel and blocks AP's valid CSA element (jamming-based MitM). \\
\hline
Spoofing & Sends a fake CSA announcing channel switch to $y$ to STA and forces AP to switch to channel $z$ (MitM). \\
\hline
Spoofing & Adversary injects a spoofed frame element with original preamble bits. \\
\hline
Spoofing & Adversary injects a frame with spoofed preamble bits. \\
\hline
Replay & Adversary replays the chain of preamble bits (i.e., replays a signature). \\
\hline
\rowcolor{gray!30} 
\multicolumn{2}{|c|}{\textbf{\circled{B} Cryptographic protocol verifier (CPV) -- Components verified}} \\
\hline
\rowcolor{gray!15} 
\multicolumn{2}{|c|}{\textit{\textbf{Asymmetric Approach}}} \\
\hline
\rowcolor{gray!7} 
\textbf{Properties} & \textbf{Automatic end-to-end verification} \\
\hline
Secrecy & Private key of AP \\
\hline
Authentication & Digital signature \\
\hline
Integrity & AP's MAC address and timestamp \\
\hline
Integrity & Operating channel, frame sequence number \\
\hline
\rowcolor{gray!15} 
\multicolumn{2}{|c|}{\textit{\textbf{Symmetric Approach}}} \\
\hline
\rowcolor{gray!7} 
\textbf{Properties} & \textbf{Automatic end-to-end verification} \\
\hline
Secrecy & Symmetric key of AP-STA pair \\
\hline
Authentication & HMAC \\
\hline
Integrity & AP's MAC address and timestamp \\
\hline
Integrity & Operating channel, frame sequence number \\
\hline
\end{tabular}
\end{table}

\subsubsection{Impact of Error-correction}\label{error_coding}

We utilized MATLAB's WLAN Toolbox to see the impact of error-correcting coding on the preamble bits (e.g., digital signature slices) of IEEE 802.11ax signals on \gls*{ber} and \gls{sr}. Due to constraints in modifying the firmware of commercially available APs, our experiments could not include the implementation coding schemes directly on the hardware. Instead, we focused on a simulation-based approach using MATLAB's WLAN Toolbox. To simplify our simulation, we assumed that the signature bits embedded in the preamble are constant. The simulation involved transmitting a signature, divided into $\mathcal{N}$ slices (corresponding to $\mathcal{N}$ pre-authentication frames from the AP), over two distinct channels: AWGN and Rayleigh. We considered the following $\mathcal{N}$ values: $\{13, 14, 15\}$ and the results show the average \gls*{ber} and \gls*{sr} across the $\mathcal{N}$ values.

Our results, as depicted in~\figurename~\ref{ber_coded} and \ref{sr_coded}, demonstrate the significant impact of applying error-correction coding on both \gls*{awgn} and Rayleigh channels. In the AWGN channel, without coding, the \gls*{ber} dropped from $15.86\%$ to virtually $0$ at $12$ dB SNR. However, with coding, this improvement is markedly accelerated, achieving virtually $0$ \gls*{ber} as early as $6$ dB SNR. Similarly, the \gls*{sr} in the AWGN channel increased from $3.74\%$ to $97.36\%$ at $12$ dB SNR without coding, whereas with coding, it reached $100\%$ at the same SNR level, indicating a complete success in transmission under these conditions. In Rayleigh channels, the impact of coding is even more pronounced. While the uncoded \gls*{ber} dropped from $28.52\%$ to $1.23\%$ at $12$ dB SNR, coded signals improved from $49.42\%$ to just $0.01\%$ at the same SNR level. Similarly, the \gls*{sr} improved from $34.40\%$ to $93.4\%$ at $12$ dB SNR, and further to $96.91\%$ at $15$ dB SNR with coding, compared to a maximum of $88.67\%$ without coding.

\subsection{Formal Security Verification}\label{formalproposed}
To systematically verify the correctness and security of our proposed protocol, we use a combination of \gls*{mc} and \gls*{cpv}. To model a protocol with PHY-layer scenarios (e.g., channel switch, jamming a frame, etc.), \gls*{mc} is an appropriate choice, specifically for inspecting whether the model meets the temporal trace property to achieve correctness. Our protocol also uses cryptography (hash, digital signature), hence, it is important to verify its cryptographic aspects, such as message integrity, authentication of a transmitter, etc. using a CPV.

\subsubsection{Symbolic Model Checking}

We developed an abstract model, $\mathcal{M}$, to represent our proposed solution, extending the Wi-Fi CE phase framework from~\cite{hoque2022systematically}. This model comprises two finite state machines (FSMs)—one each for the station (STA) and AP—defined by a set of three states, as shown in the right side of Table~\ref{formalmodel}, $\alpha = \{\alpha_1, \alpha_2, \alpha_3\}$, representing `\textit{disconnected}', `\textit{CE}', and `\textit{connected}', respectively. The FSM dynamics are driven by transitions ($\tau$), each defined by a condition-action pair, where the condition is based on the latest frame transmitted or received, and the action, which can be null, triggers the transition. Specifically, the model delineates key state transitions and introduces new variables to track channel, time constraint, and frame sequence numbers, among others. 

The simplified transitions at a station FSM are listed in Table~\ref{formalmodel}. For example, the first transition $\tau_1$ is from state $\alpha_1$ to $\alpha_2$, i.e., from \textit{disconnected}' to \textit{CE}' when condition ${\mathcal{C}^{as1}}$ is met. ${\mathcal{C}^{as1}}$ signifies that a station receives the first pre-authentication frame from an AP. In this context, $\mathcal{C}$ represents a CE frame transmission, `$as$' denotes the frame direction from AP to station, and `$1$' indicates the first pre-authentication frame. Similarly, at the AP FSM, the $\tau_1$ is ${\mathcal{C}^{as1}}$, indicating it enters the \textit{CE} state when it sends the first pre-authentication frame to a station. The other transitions occur successfully if and only if the conditions are met—as shown in Table~\ref{formalmodel}, these conditions include the correct operating channel number ($\phi_{Ch}$), frame sequence number ($\phi_{Sq}$), a pre-authentication frame containing the correct signature slice ($f_s$), a frame received within the time-bound range ($t_{in}$), and if the received frame is within its the temporal limit ($\mathcal{L}$).

The adversary model, $\mathcal{M}_{adv}$, simulates potential attacks within $\mathcal{M}$'s framework. Using the symbolic model checker NuSMV~\cite{nusmv}, we evaluated $\mathcal{M}_{adv}$ for vulnerabilities by checking all possible executions against predefined adversary actions (see Table~\ref{formalverification}-\circled{A}). The analysis did not yield any counterexamples, indicating the absence of system vulnerabilities under the tested conditions.

\subsubsection{Cryptographic Protocol Verification} We employed ProVerif~\cite{proverif}, an advanced tool for automatically verifying cryptographic protocols, to validate the cryptographic security properties of our proposed scheme comprehensively. ProVerif evaluates the scheme against a series of security criteria, marking a component's verification as failed with a FALSE outcome. ProVerif simulates all conceivable adversary actions against our proposed technique's security mechanisms, including attacks on the secrecy of an AP's private key, the integrity of message hashing, and the authenticity of digital signatures and HMAC (listed in Table~\ref{formalverification}-\circled{B}). ProVerif analysis affirmed the security of each component by returning TRUE across all checks, indicating that our scheme successfully meets the security properties.

%% file: 6.related.tex
\section{Related Work}\label{relatedwork}

\noindent\textit{AP authentication. }There are different approaches to authenticate an AP's legitimacy, such as fingerprinting using \gls*{cfo}~\cite{hua2018accurate}, phase errors between subcarriers~\cite{liu2019realtime}, the correlation between the \gls*{rss} and the transmitter’s location~\cite{sheng2008detecting}, etc. However, any changes in the transmitter hardware, channel, or configuration can significantly impact the accuracy of such techniques (even if those features are unclonable, which is not often the case). Also, this technique cannot verify the operating channel. Alternatively, traffic analysis~\cite{watkins2007apassive} or the packet characteristics in the time domain~\cite{han2011atimingbased} can reveal the presence of a rogue AP, but only after it is already connected to the stations. 

\noindent\textit{Protecting CE in Cellular Networks.} In~\cite{singla2021look}, an ID-based signature was suggested to protect the 5G \gls*{ce} against fake base stations (BS). In~\cite{hussain2019insecure}, a PKI-based solution for the 4G networks was proposed to authenticate a BS by user equipment during connection establishment. The broadcast signals from any BS were digitally signed with its private key and SIM cards equipped with certificates (a reasonable assumption in cellular networks). This approach cannot be used in WLANs because we cannot assume every Wi-Fi device is pre\hyp{}loaded with the certificate of every AP. This solution is not feasible for a Wi-Fi network's \gls*{ce} frames, as Wi-Fi devices do not have SIM cards. Therefore, it would require high maintenance, additional storage, and \gls*{ce} time, and communication costs. 

\noindent\textit{Bounding protocols. }Several bounding protocols have been proposed to prevent relay attacks. One protocol measures the round-trip time of a challenge-response interaction between the user and verifier devices, presented in~\cite{brands1994distance}, while another utilizes radio frequency identification (RFID) technology to verify physical proximity, presented in~\cite{hancke2005anrfid}. In~\cite{rasmussen2010realization}, the feasibility of deploying distance-bounding protocols was demonstrated in real-world applications. Our work is mainly centered on the investigation of the frame element modification time required by an adversary, in addition to the time taken by the frames to travel the distances between the AP and the station, as well as the (frame) processing times.

%\noindent\textit{Formal Security Methods to Analyze Wireless Protocols. }In~\cite{eian2012aformal}, formal methods were applied to discover some DoS attacks (specifically, protocol deadlock vulnerabilities) in the IEEE 802.11w protocol %(that only supports the frames that are exchanged after the key-generation stage) ~\cite{802.11i}. %Therefore, no security support for the pre\hyp{}authentication frames are there in the standard. Without analyzing Wi-Fi pre\hyp{}authentication's different stages that comes before the four-way handshake (network discovery, open authentication, association), it cannot be claimed to be \textit{secure}. Three critical procedures (attach, detach, and paging at the non-access stratum protocol layer) of 4G/LTE were formally analyzed to check their implementations in cellular devices for noncompliance with the standard, and the 5G non-access stratum and radio resource control layer in~\cite{hussain2018lteinspector}, \cite{hussain2021noncompliance}, and~\cite{hussain20195greasoner}, respectively.

%% file: 7.conclusion.tex
\section{Conclusion}\label{conclusion}

Reflecting on the escalating threats targeting the pre-authentication phase of Wi-Fi networks, we presented a novel defense mechanism. By integrating AP authentication at both the PHY and MAC layers, we effectively mitigate relay and spoofing attacks. Our approach utilizes extensive simulations and real-world experiments, including the deployment of a USRP transceiver designed for minimal-latency frame capture and relay. The results demonstrated over $97.5\%$ accuracy in relay attack detection and affirmed the practicality of our method with minimal delay impact. 